\documentclass[structabstract]{aa}
\usepackage{graphicx}
\usepackage{txfonts}
\usepackage{natbib}
\usepackage{longtable,lscape}
\usepackage{color}
\begin{document}
   \title{Chemical abundance analysis of the open clusters Berkeley 32, NGC 752, Hyades, and Praesepe\fnmsep\thanks{Based on observations collected with the fiber spectrograph FOCES at the 2.2m Calar Alto Telescope. Also based on data from 2MASS survey and the WEBDA, VALD, NIST, and GEISA online database.}}

\titlerunning{Abundances of five Open Clusters} 

   \author{R. Carrera\inst{1,2,3,4} \and E. Pancino\inst{3}}

   \authorrunning{Carrera et al.}
   \institute{Instituto de Astrof\'{\i}sica de Canarias, La Laguna, Tenerife, Spain\\
             \email{rcarrera@iac.es}
             \and
             Departamento de Astrof\'{\i}sica, Universidad de La Laguna, Tenerife, Spain\\
             \and
             INAF-Osservatorio Astronomico di Bologna, Bologna, Italy\\
             \and
	     Centro de Investigaciones de Astronom\'{\i}a, M\'erida, Venezuela
             }

   \date{Received September 15, 2110; accepted March 16, 2220}

 
  \abstract
   {Open clusters are ideal test particles for studying the chemical evolution of the
   Galactic disc. However, the number and accuracy of existing high-resolution abundance determinations, not
   only of [Fe/H], but also of other key elements, remains largely insufficient.}
   {We attempt to increase the number of Galactic open clusters that have high quality abundance
   determinations, and to gather all the literature determinations published so
   far.}
   {Using high-resolution (R$\sim$30000), high-quality (S/N$\geq$60 per pixel), we
   obtained spectra for twelve stars in four open clusters with the fibre
   spectrograph FOCES, at the 2.2 Calar Alto Telescope in Spain. We employ a classical equivalent-width analysis to obtain accurate abundances of sixteen
   elements: Al, Ba, Ca, Co, Cr, Fe, La, Mg, Na, Nd, Ni, Sc, Si, Ti, V, and Y. 
   We derived oxygen abundances derived by means of spectral synthesis of the 
   6300~\AA\  forbidden line.}
   {We provide the first determination of abundance ratios other than Fe for
   NGC~752 giants, and ratios in agreement with the literature for the Hyades, Praesepe,
   and Be~32. We use a compilation of literature data to study Galactic trends of
   [Fe/H] and [$\alpha$/Fe] with Galactocentric radius, age, and height above the
   Galactic plane. We find no significant trends, but some indication for a
   flattening of [Fe/H] at large $R_{gc}$, and for younger ages in the inner
   disc. We also detect a possible decrease in [Fe/H] with $|$z$|$ in the outer
   disc, and a weak increase in [$\alpha$/Fe] with $R_{gc}$.}
   {}

   \keywords{Stars: abundances -- Galaxy: disc -- Galaxy: open clusters and
   associations: individual: NGC~752; Hyades; Berkeley~32; Preasepe (M~44)}

   \maketitle
%

\section{Introduction}\label{sec1}

Open clusters (hereafter OC) are ideal {\em test particles} for studying the evolution
of metallicity with time, inferring the so-called {\em age-metallicity relation}, and with
Galactocentric radius, the {\em metallicity gradient}, measuring in the Galactic disc.
Their properties can be determined with smaller uncertainties than for field stars, since they are coeval group of stars at the same distance that have a
homogeneous chemical composition. Unfortunately, of the $\simeq$1700 known OC
\citep[e.g.][]{dias2002}, only $\simeq$140 possess some metallicity
determination, mostly obtained from photometric indicators, such as Washington or
Str\"omgren photometry \citep[see][and references therein]{twarog1997,chen2003}
and low-resolution spectroscopy \citep[e.g.][]{friel1993,friel02}.

The most accurate way to determine the chemical abundances is to analyse 
high-resolution spectroscopy. It allows us to investigate not only metallicity,
but also abundance ratios -- with respect to iron or hydrogen -- of other chemical
species such as $\alpha$-elements, $s$-process elements, and $r$-process
elements, which are synthesised in different environments and on different
timescales (e.g. SNe Ia, SNe II, giants, supergiants, etc). In the past few years, a number of research groups have addressed the challenge of increasing the number of OC
with chemical abundances determined from high-resolution spectroscopy
\citep[e.g.][]{sestito2004,dorazi2006,sestito2006,bragaglia2008,pace2008,dorazi2009,friel2010,pace2010,pancino2010a,jacobson2011}.
However, the number of OC with chemical abundances determined with this technique
is still small (see Section \ref{sec6}), and significant uncertainties remain in
the determinations of both the metallicity gradient and the age-metallicity relation,
which are the fundamental ingredients of chemical evolution models.

\begin{table*}
\begin{minipage}[t]{17.5cm}
\caption{Observing logs and programme star properties.}
\label{obslog}
\centering          
\renewcommand{\footnoterule}{}  
\begin{tabular}{l c c c c c c c c c c c c} 
\hline\hline
	Cluster & Star & $\alpha_{2000}$ & $\delta_{2000}$ &  B & V & R  &
        I\footnote{All I magnitudes are in the Johnson system (I$_J$) with the exception of
        those of the Be~32 stars which are in the Cousins system (I$_C$).} & K$_{S}$ & n$_{exp}$ & t$^{tot}_{exp}$ & S/N$^{tot}$\\
	 &  & (hrs) & (deg) & (mag) & (mag) &(mag) & (mag)& (mag)&  & (sec)& \\
            \hline
	Be 32\footnote{Star names from \citet{Richtler2001};
	Coordinates, B, V \& I$_C$ magnitudes from \citet{dorazi2006};
	K$_S$ magnitudes from 2MASS.}
	  & 0456 & 06:58:08.2 & +06:24:19.6 & 14.76 & 13.67 & --- & 12.53 & 11.03 & 7 & 18900 & 60 \\
	  & 1948 & 06:58:04.2 & +06:27:17.1 & 14.50 & 13.37 & --- & 12.20 & 10.68 & 6 & 16200 & 70 \\
	NGC 752\footnote{Star names from \citet{Heinemann1926}; 
	Coordinates from \citet{hog2000}; 
	B \& V magnitudes from \citet{jennens75}; 
	K$_S$ magnitudes from 2MASS.}
	& 001 & 01:55:12.6 & +37:50:14.6 & 10.47 & 9.51 & --- & --- & 7.23 & 4 & 2400 & 160 \\
	& 208 & 01:57:37.6 & +37:39:38.1 & 10.04 & 8.97 & ---  & --- & 6.41 & 4 & 2400 & 180 \\
	 & 213 & 01:57:38.9 & +37:46:12.5 & 10.08 & 9.07 & --- & --- & 6.68 & 3 & 1800 & 80 \\	 
	 & 311 &01:58:52.9 & +37:48:57.3 & 10.11 & 9.07 & --- & --- & 6.64 & 4 & 2400 & 100 \\
	Hyades\footnote{Star names from \citet{vanBueren1952}; 
	Coordinates from \citet{perryman1997}; 
	B, V, R \& I$_J$ magnitudes from \citet{Johnson1966}; 
	K$_S$ magnitudes from 2MASS.}
	                & 028 ($\gamma$ tau) & 04:19:47.6 & +15:37:39.5 & 4.64 & 3.65 & 2.92 & 2.45 & 1.52 & 2 & 120 & 560 \\
	 (Mel~25) & 041 ($\delta$ tau) & 04:22:56.1 & +17:32:33.0 & 4.75 & 3.76 & 3.03 & 2.56 & 1.64  & 3 & 180 & 450 \\
	 & 070 ($\epsilon$ tau) & 04:28:37.0 & +19:10:49.5 & 4.55 & 3.54 & 2.81& 2.31 & 1.42  & 3 & 180 & 270 \\
	Praesepe \footnote{Star names from \citet{kleinWassink1927}; 
	Coordinates from \citet{perryman1997}; 
	B, V \& R magnitudes from \citet{coleman82}; I$_J$ magnitudes from \citet{mendoza1967,Johnson1966},
	K$_S$ magnitudes from 2MASS.}
	          & 212 & 08:39:50.7 & +19:32:27.0 & 7.53 & 6.58 & 5.87 & 5.38 & 4.39 & 4 & 240 & 165 \\
	(NGC~2632) & 253 & 08:40:06.4 & +20:00:28.1 & 7.35 & 6.38 &5.67 & 5.20 & 4.20 & 4 & 240 & 215 \\
	(M~44)   & 283 & 08:40:22.1 & +19:40:11.9 & 7.42 & 6.41 & 5.68 & 5.21 & 4.18 & 2 & 120 & 150 \\
	\hline\hline
\end{tabular}
\end{minipage}
\end{table*}

In this paper, the second of a series initiated by \citet[hereafter Paper
I]{pancino2010a}, we present high quality and homogeneous measurements of chemical
abundances for red clump stars in four OC: Be~32, NGC~752, Hyades, and Praesepe.
The Hyades is the nearest OC and its four known red giants have been widely
studied
\citep{schuler2009,mishenina2007,fulbright2007,schuler2006,mishenina2006,boyarchuk2000,luck1995},
hence it provides a very good reference frame to compare our abundances with
the literature. Both NGC~752 and Praesepe have been well-studied, but all information about
their chemical composition is based mainly on their main-sequence stars
\citep[e.g.][]{pace2008,an2007,sestito2004,burkhart1998,hobbs1992}. To our
knowledge, there have been no recent measurements of the chemical abundance of their giants
from high-resolution spectroscopy. Finally, Be~32 has been the subject of some studies
\citep[e.g.][]{Richtler2001,friel2010,bragaglia2008,dorazi2006}.
The properties and previous studies of each cluster is described in more depth
in Section \ref{sec5}.

This paper is structured as follows: observations and data reduction are
described in Section \ref{sec2}; equivalent-width measurements are presented in
Section \ref{sec3}, together with the abundance analysis and its uncertainties;
results are compared with the literature in Sections \ref{sec5}, \ref{sec6}, and
\ref{sec7}; and finally our main conclusions are summarised in Section \ref{sec8}.

\section{Observational material}\label{sec2}

\begin{figure}
\centering
\includegraphics[height=16cm,width=\columnwidth]{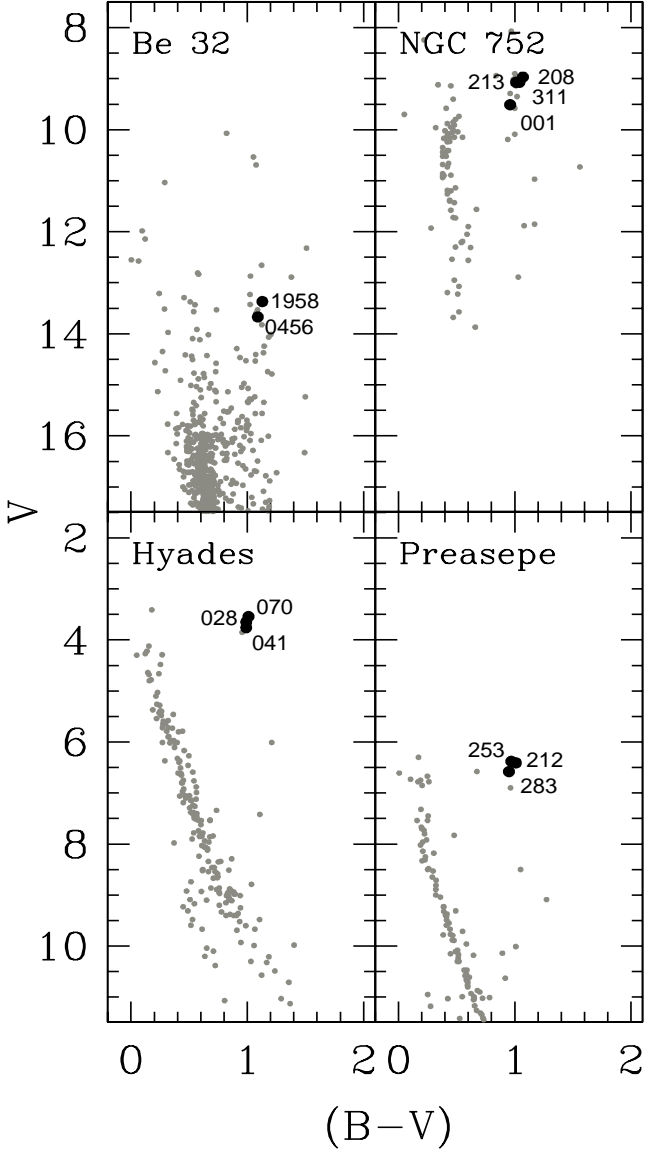}
\caption{Location of target stars (large black dots with star ID labels) in the
color-magnitude diagrams of their respective parent clusters (small grey dots).}
\label{CMDs}%
\end{figure}

A total of twelve stars spread in the four OC were observed. They were selected
from the WEBDA\footnote{{\tt http://www.univie.ac.at/webda}} database
\citep{mermilliod1995}, and the 2MASS\footnote{{\tt
http://www.ipac.caltech.edu/2mass}. 2MASS (Two Mi\-cron All Sky Survey) is a
joint project of the University of Massachusetts and the Infrared Processing and
Analysis Center/California Institute of Technology, funded by the National
Aeronautics and Space Administration and the National  Science Foundation.}
survey \citep{2mass,2mass2}. Table \ref{obslog} summarizes the identifications,
coordinates, and magnitudes of each target star. Their position in the
color-magnitude diagram taken from \citet{dorazi2006}, \citet{johnson1953}, \citet{johnson1955}, and \citet{johnson1952} for Berkeley~32, NGC~752, Hyades, and Preasepe, respectively, are
shown in Figure \ref{CMDs}.

Observations were carried out with the fibre echelle spectrograph FOCES
\citep{foces} attached at the 2.2~m Calar Alto Telescope (Almeria, Spain) between
the 1 and 3 of January 2005. The chosen set-up provides a spectral resolution
(R=$\lambda/\delta\lambda$) of about 30\,000. In summary, all stars were observed
in 2--7 exposures lasting 10--30 min each, depending on their magnitudes, until a
global signal-to-noise ratio (S/N) of at least 60 per pixel was reached around
6000 \AA. Exposures with S/N$<$20 were neglected because they were too
noisy. Finally, sky exposures as long as our longest exposures (30 min) were
taken, but the levels were sufficiently low for us to avoid sky subtraction (as in
Paper~I). The number of useful exposures, the total integration time, and the
global S/N for each star are listed in the last three columns of Table
\ref{obslog}. 

\subsection{Data reduction}

Various steps of data reduction were performed exactly as in Paper I. Briefly, the frames were
de-trended with the IRAF\footnote{Image Reduction and Analysis Facility, IRAF is
distributed by the National Optical Astronomy Observatories, which are operated
by the Association of Universities for Research in Astronomy, Inc., under
cooperative agreement with the National Science Foundation.} tasks
\textit{ccdproc} and \textit{apflatten}. The spectra were then extracted, wavelength-calibrated, normalized, and the echelle orders were merged using tasks in the
IRAF \textit{echelle} package. Finally, the noisy ends of each combined spectrum
were cut, allowing for an effective wavelength coverage from 5000 to 9000 \AA.

Before combining all exposures of each star, we removed sky absorption features
(telluric bands of O$_2$ and H$_2$O) with the help of the IRAF task
\textit{telluric}. The same two hot, rapidly rotating stars, HR 3982 and HR 8762,
of Paper I were used. The strong O$_2$ band around 7600 \AA\ had been saturated and
therefore could not be properly removed. This spectral region was not
used for the abundance analysis, in addition to the small gaps between echelle
orders that appeared after $\lambda\simeq$8400 \AA.

\subsection{Radial velocities}

\begin{table}
\begin{minipage}[htb]{\columnwidth}
\caption{Heliocentric radial velocity measurements and 1$\sigma$ errors
($V_r \pm \delta V_r$)$_{here}$ for each programme star. Literature
measurements are also reported with their uncertainties
($V_r \pm \delta V_r$)$_{lit}$.}             
\label{radvel}      
\centering          
\renewcommand{\footnoterule}{}  
\begin{tabular}{l l c c c c}     
\hline\hline       
Cluster  & Star & $(V_r\pm\delta V_r)_{here}$ & $(V_r\pm\sigma V_r)_{lit}$ \\
         &      &(km~s$^{-1}$)       &(km~s$^{-1}$) \\
\hline
Be~32\footnote{\citet{dorazi2006}.}      
	 & 0456 & 105.59$\pm$0.54 & 110.0$\pm$1.2\\
	 & 1948 & 104.78$\pm$0.35 & 105.5$\pm$4.9 \\
NGC~752\footnote{\citet{mermilliod98}.}
	 & 001 & 5.49$\pm$0.44 & 4.79$\pm$0.15 \\
         & 208\footnote{Spectroscopic binary according to \citet{pourbaix2004}.} & 1.10$\pm$0.23 & 4.86$\pm$0.06 \\
         & 213 & 5.11$\pm$0.42 & 5.50$\pm$0.10 \\
         & 311 & 6.00$\pm$0.30 & 5.28$\pm$0.08 \\
Hyades\footnote{\citet{griffin88}.}
         & 028 & 38.15$\pm$0.43 & 39.28$\pm$0.12 \\
         & 041 & 38.56$\pm$0.36 & 39.65$\pm$0.08 \\
         & 070 & 38.26$\pm$0.35 & 39.37$\pm$0.07 \\
Praesepe\footnote{\citet{Famaey2005}.}
(NGC~2632) & 212 & 35.96$\pm$0.36 & 34.81$\pm$0.21 \\
         & 253 & 34.39$\pm$0.27 & 33.67$\pm$0.22 \\
         & 283 & 34.67$\pm$0.39 & 34.35$\pm$0.20 \\
\hline
\end{tabular}
\end{minipage}
\end{table}

We used DAOSPEC \citep{daospec} to measure the observed radial velocities for each
individual exposure with S/N$\geq$20, using $\simeq$ 300 absorption lines of
different elements, with typical uncertainties of about 0.1 km s$^{-1}$ (see
Paper I for details). We used the same linelist as the one used for abundance
determinations (see Section \ref{sec3} for details). Heliocentric corrections
were obtained with the IRAF task {\sl rvcorrect}, with a negligible uncertainty
of smaller than 0.005 km s$^{-1}$. We also used DAOSPEC to determine the absolute
zero-point of the radial velocity determinations, using a list of telluric
absorption lines as the input linelist, obtained from the GEISA
database \citep{geisa1,geisa2}. The resulting zero-point corrections, based on
$\simeq$250 telluric lines, are generally no larger than  $\pm$1 km s$^{-1}$,
with a typical error of about $\simeq$0.5 km s$^{-1}$.

The final values, computed as the weighted mean of heliocentric velocities
resulting from each exposure of the same star, are listed in Table~\ref{radvel}.
Our determinations are generally in close agreement with literature values to within
3$\sigma$, except for star 208 in NGC~752, which has a slightly smaller44 radial
velocity than other objects in this cluster. The fact that this star was
recognised as a spectroscopic binary \citep[see][]{pourbaix2004,mermilliod2007}
explains the disagreement. According to its radial velocity curve \citep{mermilliod2007}, we observed this binary near minimum, which implies that we observed
only one of the components of the system. For this reason, and because derived
abundances are in good agreement with those of other stars in the same cluster,
we retained this object in our final sample. In summary, we considered all the
observed targets as likely members of their respective clusters. 

\subsection{Photometric parameters}\label{secphot}

First guesses of the atmospheric parameters effective temperature
($T_{\rm{eff}}$), logarithmic gravity ($\log g$), and microturbulent velocity
($v_t$), for our target stars were derived from a photometric data listed in
Table~\ref{obslog}, as described in Paper~I. In brief, $T_{\rm{eff}}$ were
obtained using the \citet{alonso99} and \citet{montegriffo98} colour-temperature
relations, both theoretical and empirical, and the dereddened colours (B-V)$_0$,
(V-I$_J$)$_0$, (V-R)$_0$, and (V-K$_S$)$_0$. We assumed the E(B-V) values listed
in Table \ref{phot} and the reddening laws of \citet{cardelli89}. In the case of Be~32, we have I$_C$ magnitudes instead of I$_J$ ones, so we dereddened
(V-I$_C$) with the law of \citet{dean1978}, and converted it into (V-I$_J$)$_0$
with the transformations by \citet{bessell1979}. The 1$\sigma$ errors in each
$T_{\rm{eff}}$ estimate were computed using the magnitude and reddening
uncertainties together with the standard deviation in the colour-temperature
relationships used. The photometric $T_{\rm{eff}}$ estimates, listed in Table
\ref{pars}, are the weighted mean of the different values obtained from each
considered colour and colour-temperature relations.

Photometric gravity estimates were derived from the above $T_{\rm{eff}}$ and the
bolometric corrections, BC$_V$, derived using the \citet{alonso99} prescriptions
and the fundamental relationships

\begin{eqnarray}
\nonumber
\log \frac{g}{g_\odot}=\log \frac{M}{M_\odot}+2\log
\frac{R_\odot}{R},\\
\nonumber
0.4 (M_{\rm{bol}}-M_{\rm{bol},\odot})=-4\log
\frac{T_{\rm{eff}}}{T_{\rm{eff},\odot}}+2\log \frac{R_\odot}{R},
\end{eqnarray}

\noindent where red clump masses, listed in the last column of Table \ref{pars}, were
extrapolated from Table 1 of \citet{girardi2001}. We assumed that $\log
g_{\odot}=4.437$, $T_{\rm{eff},\odot}=5770~\rm{K}$ and $M_{\rm{bol},\odot}=4.75$,
in conformity with the IAU recommendations \citep{andersen1999}. As above, we
averaged all our estimates to obtain $\log g^{(phot)}$, listed in column 5 of
Table~\ref{pars}.

As discussed in Paper I, the photometric estimate of the microturbulent
velocity, $v_t$, was obtained using the prescriptions both of \citet{ramirez2003},
$v_t=4.08-5.01~10^{-4}~T_{\rm{eff}}$, and of \citet{carretta2004},
$v_t=1.5-0.13~\log g$. The latter velocity, which takes into account the effect described
by \citet{magain1984}\footnote{However, see the discussion by \citet{mucciarelli11}
about the pros and cons of the \citet{magain1984} correction, which depends
heavily on data quality and line selection effects.}, is on average lower by
$\Delta v_t=0.50 \pm 0.03$~km~s~$^{-1}$ than the \citet{ramirez2003} estimate.
Therefore, we chose not to average the two estimates, but to use them as an
indication of the $v_t$ range to explore in our abundance analysis (see
Section~\ref{sec4}).

\begin{table}
\begin{minipage}[htb]{\columnwidth}
\caption{Adopted cluster parameters. When more than one determination
exists, the average is shown with 1~$\sigma$ errors.}
\label{phot}
\centering
\renewcommand{\footnoterule}{}  
\begin{tabular}{l c c c }     
\hline\hline       
Cluster   & E(B--V) & (m-M)$_o$ & Age \\
          & (mag)   & (mag)     & (Gyr) \\
\hline     	     
Be~32\footnote{Averages of measurements by \citet{kaluzny1991}, \citet{carraro1994}, \citet{dutra2000}, \citet{Richtler2001}, \citet{tadross2001}, \citet{lata2002}, \citet{salaris2004}, \citet{dorazi2006}, and \citet{tosi2007}.}     
          & 0.15$\pm$0.05 & 12.62$\pm$0.18 & 4.8$\pm$1.5 \\
NGC~752\footnote{Averages of measurements by \citet{johnson1953},
      \citet{roman1955}, \citet{johnson1961}, \citet{rohlfs1962}, \citet{arp1962}, \citet{eggen1963}, \citet{crawford1970},  \citet{patenaude1978}, \citet{hardy1979}, \citet{nicolet1981}, \citet{twarog1983}, \citet{barry1987}, \citet{nissen1988}, \citet{mazzei1988}, \citet{eggen1989}, \citet{francic1989}, \citet{boesgaard1991}, \citet{dzervitis1993}, \citet{carraro1993},  \citet{meynet1993}, \citet{daniel1994}, \citet{piatti1995},  \citet{dinescu1995},  \citet{milone1995}, \citet{claria1996}, \citet{dutra2000}, \citet{loktin2001}, \citet{blake2002}, \citet{blake2004}, \citet{salaris2004}, \citet{bartasiute2007},   \citet{taylor2007}, and \citet{giardino2008}.}
          & 0.038$\pm$0.002\ & 8.04$\pm$0.23 & 1.59$\pm$0.45\\	 
Hyades
          & $\leq$0.001\footnote{Derived by \citet{taylor2006} from a review of published values.} & 3.34$\pm$0.01\footnote{Averages of measurements obtained from the Hipparcos parallaxes by \citet{pinsonneault1998}, \citet{perryman1998}, \citet{narayanan1999}, \citet{loktin2001}, and \citet{percival2003}.} & 0.70$\pm$0.07\footnote{Averages of measurements by \citet{eggen1998}, \citet{loktin2001}, \citet{salaris2004}, \citet{jameson2008}, and \citet{bouvier2008}.} \\
Praesepe
          &0.027$\pm$0.004$^c$  & 6.22$\pm$0.02\footnote{Averages of measurements obtained from the Hipparcos parallaxes by \citet{pinsonneault1998}, \citet{perryman1998}, \citet{vanleeuwen1999}, \citet{loktin2000}, \citet{loktin2001}, and \citet{percival2003}.}  & 0.65\footnote{Averages of measurements obtained from the Hipparcos parallaxes by \citet{vandenheuvel1969}, \citet{maeder1971}, \citet{mathieu1988}, \citet{mazzei1988}, \citet{boesgaard1989}, \citet{tsvetkov1993}, \citet{piatti1995}, \citet{claria1996}, \citet{hernandez1998}, \citet{loktin2001}, \citet{salaris2004}, \citet{kraus2007}, and \citet{gaspar2009}}$\pm$0.25 \\
\hline                 
\end{tabular}
\end{minipage}
\end{table}

\begin{table*}
\caption{Stellar atmosphere parameters for the programme stars (see text).}
\label{pars}
\centering
\begin{tabular}{l l  c c c c c c c c}
\hline\hline
Cluster         & Star &T$_{\rm{eff}}^{(phot)}$&T$_{\rm{eff}}^{(spec)}$& $\log g^{(phot)}$&$\log g^{(spec)}$&$v_t^{(phot)}$  &$v_t^{(spec)}$  & M$_{clump}$ \\
                &      & (K)                   & (K)	               & (cgs)            & (cgs)           &(km~s$^{-1}$)   &(km~s$^{-1}$)   & (M$_\odot$) \\
\hline     	     
Be~32     & 0456 & 4759$\pm$92 & 4650 & 2.61$\pm$0.14 & 2.1 & 1.70$\pm$0.30/1.16$\pm$0.10 & 1.4 & 1.2$\pm$0.1\\
                & 1948 & 4706$\pm$99 & 4700 & 2.47$\pm$0.14 & 2.3 & 1.72$\pm$0.30/1.18$\pm$0.10 & 1.5 & 1.2$\pm$0.1\\
NGC~752 & 001 & 4949$\pm$80 & 5050 & 3.02$\pm$0.14 & 3.1 & 1.60$\pm$0.30/1.11$\pm$0.10 & 1.3 & 1.9$\pm$0.2\\
               & 208 & 4698$\pm$110 & 4600 & 2.73$\pm$0.14 & 2.9 & 1.73$\pm$0.31/1.15$\pm$0.10 & 1.2 & 1.9$\pm$0.2\\
               & 213 & 4841$\pm$86 & 4900 & 2.81$\pm$0.14 & 3.0 & 1.65$\pm$0.30/1.13$\pm$0.10 & 1.4 & 1.9$\pm$0.2\\
               & 311 & 4793$\pm$74 & 4800 & 2.80$\pm$0.14 & 3.2 & 1.68$\pm$0.30/1.14$\pm$0.10 & 1.2 & 1.9$\pm$0.2\\
Hyades    & 028 & 4865$\pm$73 & 4750 & 2.67$\pm$0.04 & 2.7 & 1.64$\pm$0.30/1.15$\pm$0.15 & 1.4 & 2.5$\pm$0.1\\
                & 041 & 4871$\pm$79 & 4800 & 2.71$\pm$0.05 & 2.8 & 1.64$\pm$0.30/1.15$\pm$0.15 & 1.4 & 2.5$\pm$0.1\\
                & 070 & 4858$\pm$95 & 4800 & 2.62$\pm$0.05 & 2.8 & 1.65$\pm$0.30/1.16$\pm$0.15 & 1.6 & 2.5$\pm$0.1\\
Praesepe & 212 & 4901$\pm$35 & 4900 & 2.70$\pm$0.07 & 2.8 & 1.62$\pm$0.30/1.15$\pm$0.14 & 1.5 & 2.6$\pm$0.3\\
  & 253 & 4869$\pm$23 & 4900 & 2.60$\pm$0.07 & 2.8 & 1.64$\pm$0.30/1.16$\pm$0.14 & 1.6 & 2.6$\pm$0.3\\
                 & 283 & 4841$\pm$29 & 4800 & 2.61$\pm$0.07 & 2.9 & 1.65$\pm$0.30/1.16$\pm$0.14 & 1.4 &  2.6$\pm$0.3\\
\hline
\end{tabular}
\end{table*}

\begin{table*}
\caption{Equivalent widths and atomic data of the programme stars. The complete version of the table is available at the CDS. Here we show a few lines to illustrate its contents.}
\label{tab-ew}      
\centering          
\begin{tabular}{c c c c c c c c c c c c c c}
\hline\hline
 & & & &\multicolumn{3}{l}{Be~32--Star~0456} &
\multicolumn{3}{c}{Be~32--Star~1948} & ... & \multicolumn{3}{c}{Praesepe--Star~283}\\ 
\hline     	     
$\lambda$ & Elem  & $\chi_{\rm{ex}}$&$\log gf$& EW  &$\delta$EW & Q & EW     &$\delta$EW & Q & ... & EW     &$\delta$EW & Q \\
(A) &  & (eV) & (dex) & (m\AA) & (m\AA) &  & (m\AA) & (m\AA) &  & ... & (m\AA) & (m\AA) & \\
\hline 
6696.79 & AL1 & 4.02 & -1.42 &  16.5 & 1.9 & 0.361 &  25.3 & 3.9 & 0.481 & ... &  30.0 & 5.0 & 0.585\\
6698.67 & AL1 & 3.14 & -1.65 &  47.8 & 2.7 & 0.432 &  43.4 & 2.6 & 0.501 & ... &  59.2 & 1.8 & 0.321\\
7361.57 & AL1 & 4.02 & -0.90 &  33.8 & 3.9 & 0.633 &  33.2 & 6.5 & 0.946 & ... &  52.1 & 2.1 & 0.173\\
7362.30 & AL1 & 4.02 & -0.75 &  48.6 & 4.3 & 0.346 &  39.7 & 2.9 & 0.737 & ... &  72.0 & 6.9 & 0.797\\
7835.31 & AL1 & 4.02 & -0.65 &  53.7 & 3.7 & 0.393 &  48.4 & 3.9 & 0.611 & ... &  95.8 & 9.8 & 1.399\\
7836.13 & AL1 & 4.02 & -0.49 &  66.0 & 7.8 & 1.279 &  68.4 & 5.1 & 0.783 & ... &  94.0 & 6.0 & 0.665\\
8772.86 & AL1 & 4.02 & -0.32 & ...   & ... & ...   &  ...  & ... & ....  & ... & 102.7 & 8.7 & 0.748\\
8773.90 & AL1 & 4.02 & -0.16 & 115.0 & 8.6 & 1.151 & 111.9 & 8.3 & 1.111 & ... & ... & ... & ...\\
5853.67 & BA2 & 0.60 & -1.00 & 119.7 & 3.9 & 1.098 & 102.0 & 2.6 & 0.343 & ... & 122.1 & 3.4 & 0.676\\
\hline
\end{tabular}
\end{table*}

\begin{table*}[!t]
\begin{minipage}[t]{\textwidth}
\caption{Abundance ratios for single cluster stars, with their 
internal and external (last column) uncertainties.}
\label{abotab}
\centering 
\begin{tabular}{l|c c c c c c |c}
\hline \hline
& \multicolumn{2}{c|}{Berkeley 32} & \multicolumn{4}{c|}{NGC~752} & External\\
Ratio         & Star 456       & Star 1948       & Star 001       & Star 208        & Star 213        & Star 311 & Uncertainty \\
\hline
$[$FeI/H$]$  & --0.33$\pm$0.02 & --0.27$\pm$0.02 & +0.07$\pm$0.01 & +0.07$\pm$0.01 & +0.04$\pm$0.01 & +0.14$\pm$0.01 &$\pm$0.03\\
$[$FeII/H$]$ & --0.30$\pm$ 0.06 & --0.29$\pm$0.06 & +0.02$\pm$0.03 & +0.06$\pm$0.03 & +0.05$\pm$0.04 & +0.18$\pm$0.12 &$\pm$0.03\\
$[$$\alpha$/Fe$]$ & --0.29$\pm$0.21 & --0.25$\pm$0.09 & +0.07$\pm$0.04 & +0.05$\pm$0.12 & +0.07$\pm$0.12 & +0.14$\pm$0.09 &$\pm$0.07\\
$[$Al/Fe$]$ & +0.15$\pm$0.06 & +0.08$\pm$0.07 & --0.11$\pm$0.04 & --0.12$\pm$0.03 & --0.06$\pm$0.06 & --0.21$\pm$0.06 &$\pm$0.05\\
$[$Ba/Fe$]$ & +0.52$\pm$0.05 & +0.35$\pm$0.17 & +0.55$\pm$0.04 & +0.52$\pm$0.04 & +0.51$\pm$0.01 & +0.57$\pm$0.06 &$\pm$0.04\\
$[$Ca/Fe$]$ & --0.06$\pm$0.08 & --0.05$\pm$0.04 & --0.02$\pm$0.03 & --0.12$\pm$0.02 & --0.09$\pm$0.03 & --0.17$\pm$0.05 &$\pm$0.06\\
$[$Co/Fe$]$ & +0.02$\pm$0.05 & +0.09$\pm$0.04 & --0.03$\pm$0.03 & +0.06$\pm$0.04 & +0.00$\pm$0.03 & +0.05$\pm$0.05 &$\pm$0.04\\
$[$Cr/Fe$]$ & --0.25$\pm$0.07 & +0.04$\pm$0.07 & +0.02$\pm$0.03 & +0.00$\pm$0.03 & --0.01$\pm$0.03 & --0.01$\pm$0.04 &$\pm$0.05\\
$[$La/Fe$]$ & --0.14$\pm$0.02 & --0.04$\pm$0.08 & +0.14$\pm$0.06 & +0.18$\pm$0.03 & +0.18$\pm$0.09 & +0.32$\pm$0.13 &$\pm$0.04\\
$[$Mg/Fe$]$ & +0.38$\pm$0.12 & +0.24$\pm$0.16 & +0.13$\pm$0.06 & +0.16$\pm$0.05 & +0.20$\pm$0.04 & +0.06$\pm$0.03 &$\pm$0.09\\
$[$Na/Fe$]$ & --0.14$\pm$0.08 & --0.08$\pm$0.10 & +0.05$\pm$0.01 & --0.07$\pm$0.02 & --0.03$\pm$0.05 &-0.10$\pm$0.05 &$\pm$0.04\\
$[$Nd/Fe$]$ & --0.05$\pm$0.13 & +0.04$\pm$0.03 & +0.29$\pm$0.14 & +0.27$\pm$0.23 & +0.34$\pm$0.11 & +0.46$\pm$0.18 &$\pm$0.13\\
$[$Ni/Fe$]$ & --0.04$\pm$0.03 & --0.01$\pm$0.03 & --0.04$\pm$0.02 & +0.00$\pm$0.02 & --0.02$\pm$0.02 & +0.03$\pm$0.03 &$\pm$0.02\\
$[$O/Fe$]$  &--0.16$\pm$0.13 & +0.15$\pm$0.11 & +0.15$\pm$0.06 & --0.06$\pm$0.05 & +0.02$\pm$0.08 & +0.00$\pm$0.06 &$\pm$0.08\\
$[$Sc/Fe$]$ & +0.02$\pm$0.05 & --0.02$\pm$0.05 & --0.02$\pm$0.05 & +0.04$\pm$0.06 & +0.05$\pm$0.06 & +0.09$\pm$0.08 &$\pm$0.05\\
$[$Si/Fe$]$ & +0.18$\pm$0.04 & +0.11$\pm$0.04 & --0.03$\pm$0.03 & +0.04$\pm$0.03 & +0.04$\pm$0.03 & +0.01$\pm$0.04 &$\pm$0.04\\
$[$TiI/Fe$]$ & --0.10$\pm$0.05 & --0.04$\pm$0.05 & +0.00$\pm$0.02 & --0.03 $\pm$0.02 & --0.08$\pm$0.02 & --0.13$\pm$0.03 &$\pm$0.03\\
$[$TiII/Fe$]$ & --0.17$\pm$0.05 & +0.01$\pm$0.07 & +0.03$\pm$0.02 & +0.08$\pm$0.07 & +0.07$\pm$0.06 & +0.15$\pm$0.13 &$\pm$0.03\\
$[$V/Fe$]$  & --0.14$\pm$0.10 & --0.07$\pm$0.05 & +0.00$\pm$0.02 & +0.16$\pm$0.05 & --0.04$\pm$0.03 & +0.05$\pm$0.06 &$\pm$0.06\\
$[$Y/Fe$]$  & --0.41$\pm$N.A. & --0.09$\pm$N.A. & --0.12$\pm$0.06 & --0.03$\pm$0.10 & +0.03$\pm$0.07 & +0.05$\pm$0.09 &$\pm$0.04\\
\hline \hline
& \multicolumn{3}{c|}{Hyades (Mel~25)} & \multicolumn{3}{c|}{Praesepe (NGC~2632)} & \\
Ratio         & Star 28        & Star 41        & Star 70  & Star 212       & Star 253	  & Star 283 & \\
\hline
$[$FeI/H$]$ & +0.12$\pm$0.01 & +0.10$\pm$0.01 & +0.11$\pm$0.01 & +0.11$\pm$0.01 & +0.18$\pm$0.01 & +0.21$\pm$0.01 &$\pm$0.03\\
$[$FeII/H$]$ & +0.13$\pm$0.03 & +0.13$\pm$0.03 & +0.09$\pm$0.03 & +0.10$\pm$0.03 & +0.16$\pm$0.03 & +0.23$\pm$0.04 &$\pm$0.03\\
$[$$\alpha$/Fe$]$ & +0.13$\pm$0.15 & +0.11$\pm$0.12 & +0.09$\pm$0.11 & +0.11$\pm$0.15 & +0.18$\pm$0.14 & +0.19$\pm$0.13 &$\pm$0.07\\
$[$Al/Fe$]$ & --0.01$\pm$0.05 & +0.00$\pm$0.05 & +0.02$\pm$0.05 & +0.01$\pm$0.04 & +0.02$\pm$0.06 & --0.04$\pm$0.05 &$\pm$0.05\\
$[$Ba/Fe$]$ & +0.37$\pm$0.05 & +0.39$\pm$0.05 & +0.31$\pm$0.05 & +0.30$\pm$0.08 & +0.27$\pm$0.06 & +0.37$\pm$0.05 &$\pm$0.04\\
$[$Ca/Fe$]$ & --0.07$\pm$0.03 & --0.06$\pm$0.03 & --0.07$\pm$0.02 & --0.07$\pm$0.02 & --0.08$\pm$0.03 & --0.11$\pm$0.03 &$\pm$0.06\\
$[$Co/Fe$]$ & +0.00$\pm$0.04 & +0.01$\pm$0.03 & +0.06$\pm$0.03 & +0.05$\pm$0.03 & +0.01$\pm$0.03 & +0.05$\pm$0.05 &$\pm$0.04\\
$[$Cr/Fe$]$ & +0.02$\pm$0.03 & +0.03$\pm$0.03 & +0.08$\pm$0.04 & +0.06$\pm$0.03 & +0.04$\pm$0.04 & +0.04$\pm$0.04 &$\pm$0.05\\
$[$La/Fe$]$ & --0.12$\pm$0.06 & --0.08$\pm$0.05 & --0.05$\pm$0.05 & --0.07$\pm$0.05 & --0.04$\pm$0.05 & --0.04$\pm$0.04 &$\pm$0.04\\
$[$Mg/Fe$]$ & +0.13$\pm$0.05 & +0.06$\pm$0.04 & +0.21$\pm$0.07 & +0.31$\pm$0.06 & +0.27$\pm$0.05 & +0.22$\pm$0.06 &$\pm$0.09\\
$[$Na/Fe$]$ & +0.19$\pm$0.02 & +0.18$\pm$0.02 & +0.18$\pm$0.02 & +0.23$\pm$0.02 & +0.30$\pm$0.03 & +0.18$\pm$0.05 &$\pm$0.04\\
$[$Nd/Fe$]$ & +0.04 $\pm$0.29 & +0.08$\pm$0.30 & +0.08$\pm$0.28 & +0.00$\pm$0.21 & +0.05$\pm$0.25 & +0.10$\pm$0.31 &$\pm$0.13\\
$[$Ni/Fe$]$ & +0.02 $\pm$0.02 & +0.04$\pm$0.02 & +0.03$\pm$0.02 & +0.01$\pm$0.02 & +0.01$\pm$0.02 & +0.04$\pm$0.03 &$\pm$0.02\\
$[$O/Fe$]$ & --0.35$\pm$0.07 & --0.25$\pm$0.05 & --0.22$\pm$ 0.07 & --0.11$\pm$0.09 & --0.14$\pm$0.07 & --0.09$\pm$0.06 &$\pm$0.08\\
$[$Sc/Fe$]$ & --0.04 $\pm$0.05 & +0.00$\pm$0.05 & --0.02$\pm$0.06 & --0.10$\pm$0.06 & +-0.03$\pm$0.05 & +0.00$\pm$0.06 &$\pm$0.05\\
$[$Si/Fe$]$ & +0.09 $\pm$0.03 & +0.09$\pm$0.02 & +0.10$\pm$0.03 & +0.06$\pm$0.03 & +0.07$\pm$0.03 & +0.04$\pm$0.03 &$\pm$0.04\\
$[$TiI/Fe$]$ & --0.12 $\pm$0.02 & --0.11$\pm$0.02 & --0.06$\pm$0.02 & --0.05$\pm$0.03 & --0.08$\pm$0.02 & --0.09$\pm$0.02 &$\pm$0.03\\
$[$TiII/Fe$]$& --0.03 $\pm$0.06 & +0.00$\pm$0.07 & --0.02$\pm$0.11 & --0.05$\pm$0.10 & --0.02$\pm$0.08 & +0.05$\pm$0.08 &$\pm$0.03\\
$[$V/Fe$]$ & +0.02 $\pm$0.04 & +0.00$\pm$0.03 & +0.09$\pm$0.03 & +0.06$\pm$0.04 & +0.04$\pm$0.03 & +0.10$\pm$0.04 &$\pm$0.06\\
$[$Y/Fe$]$ & --0.12 $\pm$0.05 & --0.06$\pm$0.06 & --0.07$\pm$0.05 & --0.11$\pm$0.10 & --0.12$\pm$0.07 & --0.11$\pm$0.09 &$\pm$0.04\\
\hline
\end{tabular}
\end{minipage} 
\end{table*}

\section{Equivalent widths and abundance analysis}\label{sec3}

We used the same linelist as that described in Paper I. In brief, all lines and
their atomic data were extracted from the VALD
 database \citep{vald}, with a few
exceptions (see Paper~I for details). Briefly, for some highly discrepant Mg
lines, we used the NIST log$gf$ values; we used the \citet{johansson03} log$gf$
for the Ni line that contaminates the [O~I] line at 6300~\AA, and provides
oxygen abundances more in line with the other $\alpha$-elements; we used the Nd
log$gf$ values by \citet{denhartog2003}, which minimize the spread in the Nd
abundance. Finally, we tried both the VALD and the NIST values for Ca, finding an
average difference of 0.17~dex (see paper~I). There is no special reason for
choosing NIST over VALD (or vice-versa), so we kept the VALD values to help maintain some
homogeneity, but we note that the Ca log$gf$ values carry a large uncertainty of
the order of 0.2~dex.

\begin{figure}
\centering
\includegraphics[width=\columnwidth]{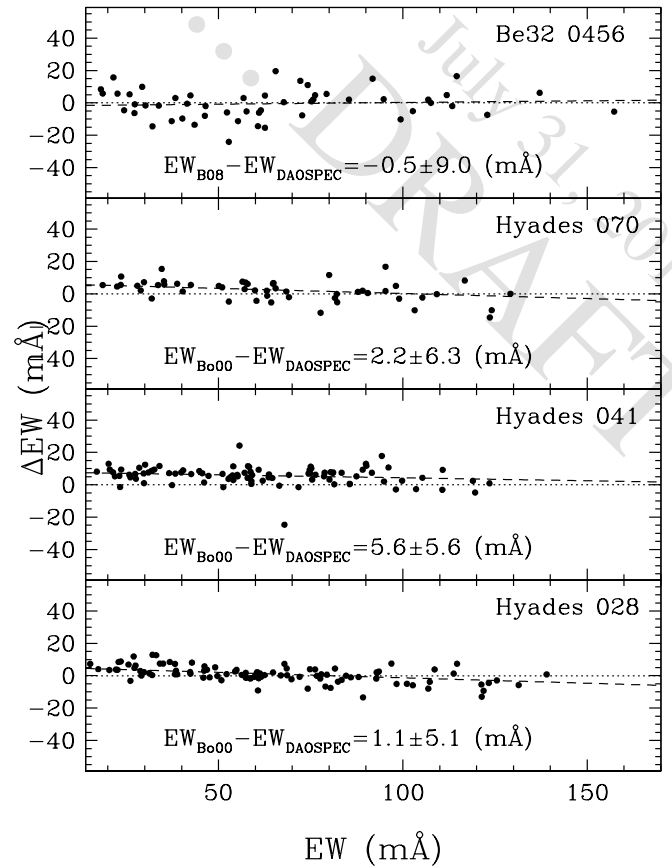}
\caption{Comparison of our EW measurements with those by \citet{bragaglia2008}
for star 0456 in Be~32, and by \citet{boyarchuk2000} for three Hyades giants.
Dotted lines mark perfect agreement (zero difference), while dashed lines are
linear fits to the data.}
\label{compew}
\end{figure}

\begin{table*}
\begin{minipage}[t]{\textwidth}

\caption{Average cluster abundances, obtained as the weighted average of the
single stars abundances in each of them.}

\label{tab-cluster}
\centering
\renewcommand{\footnoterule}{}
\begin{tabular}{l c c c c c}
\hline\hline 
Ratio           &  Be~32         & NGC~752        & Hyades        & Praesepe (M~44) \\ 
\hline
$[$Fe/H$]$      & --0.30$\pm$0.02($\pm$0.03) &  +0.08$\pm$0.04($\pm$0.03) & +0.11$\pm$0.01($\pm$0.03) & +0.16$\pm$0.05($\pm$0.03) \\
$[\alpha$/Fe$]$ & --0.04$\pm$0.14($\pm$0.07) &  +0.02$\pm$0.06($\pm$0.07) & +0.00$\pm$0.12($\pm$0.07) & +0.00$\pm$0.14($\pm$0.07) \\
\hline
$[$Al/Fe$]$  &  +0.12$\pm$0.05($\pm$0.05) & --0.12$\pm$0.06($\pm$0.05) &  +0.00$\pm$0.02($\pm$0.05) & +0.00$\pm$0.03($\pm$0.05)  \\
$[$Ba/Fe$]$  &  +0.51$\pm$0.12($\pm$0.04) &  +0.52$\pm$0.03($\pm$0.04) &  +0.36$\pm$0.04($\pm$0.04) &  +0.33$\pm$0.05($\pm$0.04)  \\
$[$Ca/Fe$]$  & --0.05$\pm$0.01($\pm$0.06) & --0.09$\pm$0.06($\pm$0.06) & --0.07$\pm$0.01($\pm$0.06) & --0.08$\pm$0.02($\pm$0.06)  \\ 
$[$Co/Fe$]$  &  +0.07$\pm$0.05($\pm$0.04) &  +0.01$\pm$0.04($\pm$0.04) &  +0.03$\pm$0.03($\pm$0.04) &  +0.04$\pm$0.02($\pm$0.04)  \\ 
$[$Cr/Fe$]$  & --0.11$\pm$0.21($\pm$0.05) &  +0.00$\pm$0.01($\pm$0.05) &  +0.04$\pm$0.03($\pm$0.05) &  +0.05$\pm$0.01($\pm$0.05)  \\ 
$[$La/Fe$]$  & --0.14$\pm$0.07($\pm$0.04) &  +0.18$\pm$0.08($\pm$0.04) & --0.08$\pm$0.04($\pm$0.04) & --0.05$\pm$0.02($\pm$0.04)  \\ 
$[$Mg/Fe$]$  &  +0.33$\pm$0.10($\pm$0.09) &  +0.12$\pm$0.06($\pm$0.09) &  +0.10$\pm$0.08($\pm$0.09) &  +0.27$\pm$0.05($\pm$0.09)  \\ 
$[$Na/Fe$]$  & --0.12$\pm$0.04($\pm$0.04) &  +0.01$\pm$0.07($\pm$0.04) &  +0.18$\pm$0.01($\pm$0.04) &  +0.25$\pm$0.06($\pm$0.04)  \\ 
$[$Nd/Fe$]$  &  +0.03$\pm$0.06($\pm$0.13) &  +0.34$\pm$0.09($\pm$0.13) &  +0.07$\pm$0.02($\pm$0.13) &  +0.04$\pm$0.05($\pm$0.13)  \\ 
$[$Ni/Fe$]$  & --0.03$\pm$0.02($\pm$0.02) & --0.01$\pm$0.03($\pm$0.02) &  +0.03$\pm$0.01($\pm$0.02) &  +0.02$\pm$0.02($\pm$0.02)  \\ 
$[$O/Fe$]$   & +0.00$\pm$0.16($\pm$0.08) & +0.03$\pm$0.04($\pm$0.08) & --0.27$\pm$0.04($\pm$0.08) & --0.11$\pm$0.03($\pm$0.08)  \\ 
$[$Sc/Fe$]$  &  +0.00$\pm$0.03($\pm$0.05) &  +0.03$\pm$0.05($\pm$0.05) & --0.02$\pm$0.02($\pm$0.05) & --0.04$\pm$0.05($\pm$0.05)  \\ 
$[$Si/Fe$]$  &  +0.14$\pm$0.05($\pm$0.04) &  +0.02$\pm$0.03($\pm$0.04) &  +0.09$\pm$0.01($\pm$0.04) &  +0.06$\pm$0.02($\pm$0.04)  \\ 
$[$Ti/Fe$]$  & --0.08$\pm$0.07($\pm$0.03) & --0.03$\pm$0.06($\pm$0.03) & --0.09$\pm$0.04($\pm$0.03) & --0.07$\pm$0.03($\pm$0.07)  \\ 
$[$V/Fe$]$   & --0.08$\pm$0.05($\pm$0.06) &  +0.01$\pm$0.09($\pm$0.06) &  +0.04$\pm$0.05($\pm$0.06) &  +0.06$\pm$0.03($\pm$0.06)  \\ 
$[$Y/Fe$]$   & --0.23$\pm$0.23($\pm$0.04) & --0.03$\pm$0.08($\pm$0.04) & --0.09$\pm$0.03($\pm$0.04) & --0.11$\pm$0.01($\pm$0.04)  \\ 
\hline
\end{tabular}
\end{minipage}
\end{table*}

\subsection{Equivalent widths with DAOSPEC}

The task DAOSPEC \citep{daospec} was used to automatically find and measure equivalent
widths (hereafter EW), by performing a Gaussian fitting of the identified lines.
DAOSPEC provides a formal error in the Gaussian fit, $\delta EW$, and a quality
parameter, $Q$ \citep[see][and Paper~I, for more details]{daospec}. The relative
error $\delta EW/EW$  and the quality parameter $Q$ can be used to distinguish
good and bad lines, and they were indeed used to select the highest quality lines for the
abundance analysis, as described in detail in Paper~I. The measured EW for our program stars are shown in the electronic version of Table \ref{tab-ew} along with the $\delta EW$ and $Q$ parameter estimated by DAOSPEC.

Four of our target stars have published EW measurements from high-resolution
spectra. These consist of three stars (namely, 028, 041, and 070) observed in the Hyades
by \citet{boyarchuk2000} with $R\sim45000$, and star 0456 in Be~32 studied by
\citet{bragaglia2008} with $R\sim40000$. We have a total of 100, 92, and 51 lines
in common for stars 028, 041, and 070 in the Hyades, respectively, and 51 lines
for star 0456 in Be~32. Figure~\ref{compew} compares the comparison between the EW
determined with DAOSPEC with the values published by \citet{bragaglia2008} and
\citet{boyarchuk2000}. The differences (see Figure~\ref{compew}) are negligible
within the uncertainties; we find a small offset of 5.6~m\AA\  in the case of star
041 in the Hyades, which is however still within 1\,$\sigma$. We can therefore
consider our measurements in good agreement with similar studies.

\subsection{Abundance analysis}\label{sec4}

Abundance calculations and spectral synthesis (for oxygen) were performed
using the latest version of the abundance calculation code originally described
by \citet{spite1967}. We used the MARCS model atmospheres developed by
\citet{edvardson1993}. We also used of ABOMAN, a tool developed by E.
Rossetti at the INAF, Bologna Observatory, Italy, which allows the semi-automatic
processing of data for several objects, using the aforementioned abundance calculation
code. The tool ABOMAN performs all the steps needed to choose the best-fit model automatically
(see below) and compute abundance ratios for all elements, and provides all
the graphical tools required to analyse the results.

The detailed procedure followed to derive the chemical abundances is described in
depth in Paper I. In brief, we calculated Fe~I and Fe~II abundances for a set of
models with parameters extending $\pm$3$\sigma$ around the photometric estimates
of Table~\ref{pars}. We chose the model that satisfied simultaneously the
following conditions: {\em (i)} the abundance of Fe~I lines should not vary with
excitation potential $\chi_{ex}$; {\em (ii)} the abundance of Fe~I lines should
not vary significantly with EW, i.e., strong and weak lines should infer the same
abundance\footnote{We decided not to use the \citet{magain1984} effect, because we
prefer to have internally consistent abundances from each line, and because of the
additional effects described by \citet{mucciarelli11}.}; {\em (iii)} the abundance
of Fe~I lines should not differ significantly from the abundance of Fe~II lines; and
{\em (iv)} the abundance of Fe~I lines should not vary significantly with
wavelength.

Once the best-fit model has been found, abundance ratios of all the measured elements
were determined, as shown in Table~\ref{abotab}, as the average of abundances
given by single lines. The internal (random) errors were then computed as $\sigma
/ \sqrt(n_{lines})$. Oxygen abundances were determined by means of spectral synthesis of
the region around the [O~I] forbidden line at 6300~\AA. In this case, the
internal uncertainty was estimated using the average abundance difference between
the best-fit spectrum and two spectra placed approximately 1$\sigma$ (of the
Poissonian noise) above and below it. Average cluster abundances
(Tables~\ref{tab-cluster}) were computed as weighted averages of abundance ratios
of single stars.

Comparison of our results with available literature is discussed in details in
Section~\ref{sec5}.

\subsection{Abundance uncertainties and the Sun}\label{sec-err}

The internal (random) uncertainty described above includes uncertainties
related to the measurement of EW and to the atomic parameters (dominated by
log$gf$ determinations). We must consider other sources of uncertainty (see
Paper~I for details) such as: the uncertainty owing to the choice of atmospheric
parameters; the uncertainty owing to the continuum normalization procedure; the
uncertainty in the reference solar abundance values.

Uncertainties due to the choice of stellar parameters were evaluated with the
method proposed by \citet{cayrel2004}. In brief, we altered the predominant
atmospheric parameter, i.e., by altering one atmospheric parameter, T$_{\rm{eff}}$, within
its uncertainty ($\sim$100 K) and re-optimizing the other parameters for the hottest
and coolest stars in our sample. We re-calculated abundances with the procedure
described in the previous Section. The external uncertainties, listed in the last
column of Table~\ref{abotab}, are estimated by averaging errors calculated with
the higher and lower temperatures for the warmest and coolest stars in our sample
(namely, stars 001 and 208 in NGC~752).

Uncertainties due to the continuum normalization procedure might also affect the
obtained EW and, therefore, the derived abundances. Their contribution is
estimated by averaging the differences between the EW obtained with the
``best-fit'' continuum and those derived by lowering and raising the continuum
level by the continuum placement uncertainty. This is calculated from Equation 7
of \citet{daospec}. The typical uncertainty caused by the continuum placement is
$\Delta EW\sim$1 m\AA\  and almost independent of the EW. This small uncertainty
has a negligible impact on the derived abundances in comparison with other sources
of uncertainty described above. Therefore, they have not been explicitly
included in the error budget.

\begin{table}
\begin{minipage}[htb]{\columnwidth}
\caption{High-resolution average Be~32 abundances.}
\label{be32} 
\centering
\renewcommand{\footnoterule}{}
\begin{tabular}{l r r r }
\hline\hline
        & {\em Here} & F10\footnote{\citet{friel2010}, from 2 stars.} &
        B08\footnote{\citet{bragaglia2008} and \citet{sestito2006}, from 10 red clump and RGB stars.} \\
\hline
R=$\lambda/\delta\lambda$ & 30000  & 28000 & 40000  \\
S/N                       & 50--100 & 100 & 45-100  \\
\hline
$[$Fe/H$]$  & --0.30$\pm$0.02 & --0.30$\pm$0.02 & --0.29$\pm$0.04  \\
$[$Al/Fe$]$ & +0.12$\pm$0.05 & +0.19$\pm$0.06 & +0.11$\pm$0.10 \\
$[$Ba/Fe$]$ & +0.51$\pm$0.12 &  --- & +0.29$\pm$0.10  \\
$[$Ca/Fe$]$ & --0.05$\pm$0.01 & --0.07$\pm$0.01 & +0.07$\pm$0.04  \\
$[$Co/Fe$]$ & +0.07$\pm$0.05 &+0.00$\pm$0.01 & --- \\
$[$Cr/Fe$]$ & --0.11$\pm$0.21 & --0.16$\pm$0.11 & --0.05$\pm$0.04 \\
$[$La/Fe$]$ & --0.14$\pm$0.07 & --- &  --- \\
$[$Mg/Fe$]$ & +0.33$\pm$0.10 & +0.13$\pm$0.01 & +0.27$\pm$0.08 \\
$[$Na/Fe$]$ & --0.12$\pm$0.04 & +0.20$\pm$0.01 & +0.13$\pm$0.02 \\
$[$Ni/Fe$]$ & --0.03$\pm$0.02 & --0.02$\pm$0.01 & +0.00$\pm$0.04 \\
$[$O/Fe$]$  & +0.00$\pm$0.16 &--0.01$\pm$0.03 & ---\\
$[$Sc/Fe$]$ & +0.00$\pm$0.03 & --- &  ---  \\
$[$Si/Fe$]$ & +0.14$\pm$0.05 &+0.27$\pm$0.05 & +0.12$\pm$0.04 \\
$[$Ti/Fe$]$ & --0.08$\pm$0.07 & --0.17$\pm$0.01 & +0.11$\pm$0.06 \\
$[$V/Fe$]$  & --0.08$\pm$0.05 & --- & --- \\
$[$Y/Fe$]$  & --0.23$\pm$0.23 & --- & --- \\
\hline
\end{tabular}
\end{minipage}
\end{table}

\begin{table*}
\begin{minipage}[t]{\textwidth}
\caption{Abundance comparison of individual Hyades stars (see text). } 
\label{tab_hyades_individual}
\centering
\renewcommand{\footnoterule}{}
\begin{tabular}{l |r r r r r r }
\hline\hline
Parameter         & Here  & S09/S06 & M07/M06 & F07 & Bo00 & LC95 \\
Resolution & 30000 & 60000 & 42000 & 30000 & 45000 & 30000  \\
S/N             & 300--600 & $\sim$500 & 100-350 & 175 & 100--300 & $>$100\\
\hline
Star &  \multicolumn{6}{c}{028 ($\gamma$ tau)}\\
\hline
T$_eff$ (K)  & 4750 & 4965 & 4955 & 4823 & 4956 & 4900\\
log$g$ (dex) & 2.7 & 2.63 & 2.7 & 2.43 & 2.83 & 2.6\\
$v_t$ (km~s$^{-1}$) & 1.4 & 1.32 & 1.4 & 1.57 & 1.35 & 2.0\\
\hline
$[$FeI/H$]$ & +0.12$\pm$0.01 & +0.14$\pm$0.08 & +0.11  & +0.16$\pm$0.05 & +0.11$\pm$0.01 & +0.13$\pm$0.02\\ 
$[$FeII/H$]$ & +0.13$\pm$0.03 & +0.22$\pm$0.16 & +0.10  & +0.09$\pm$0.08& +0.11$\pm$0.02& +0.12$\pm$0.02\\
\hline
$[$Al/Fe$]$ & -0.01$\pm$0.05 & +0.20$\pm$0.01 & ---  & +0.19$\pm$0.07 & +0.12$\pm$0.00 & --0.17$\pm$0.03\\  
$[$Ba/Fe$]$ & +0.37$\pm$0.05 & --- &--0.07& --- & +0.09$\pm$0.05 & --0.04$\pm$0.00\\ 
$[$Ca/Fe$]$ & --0.07$\pm$0.03 & --- &+0.10$\pm$0.12  & --0.01$\pm$0.11& +0.01$\pm$0.04 & --0.28$\pm$0.04\\  
$[$Co/Fe$]$ & +0.00$\pm$0.04 & --- & --- & --- & +0.02$\pm$0.02 & +0.04$\pm$0.04\\ 
$[$Cr/Fe$]$ & +0.02$\pm$0.03 & ---  & --- & --- &  --0.01$\pm$0.02 & --0.03$\pm$0.11\\ 
$[$La/Fe$]$ & --0.12$\pm$0.06 & --- & --0.23 & --- & --0.03$\pm$0.02 & --- \\ 
$[$Mg/Fe$]$ & +0.13$\pm$0.05 & +0.43$\pm$0.01 & --0.08  & +0.03$\pm$0.07 & +0.16$\pm$0.02 & +0.18$\pm$0.05\\ 
$[$Na/Fe$]$ & +0.19$\pm$0.02 & +0.49$\pm$0.05 & +0.22   & +0.05$\pm$0.11& +0.32$\pm$0.01 & +0.22$\pm$0.05\\ 
$[$Nd/Fe$]$ & +0.04$\pm$0.29 & --- & --0.15 & --- & --0.02$\pm$0.03 & +0.07$\pm$0.00\\ 
$[$Ni/Fe$]$ & +0.02$\pm$0.02 & +0.12$\pm$0.07 &--0.04$\pm$0.12  & --- & +0.00$\pm$0.03 & +0.06$\pm$0.03 \\ 
$[$O/Fe$]$ & --0.35$\pm$0.07 & --0.09$\pm$0.06 & --0.09  & --0.04$\pm$0.11 & --- & --- \\ 
$[$Sc/Fe$]$ & --0.04$\pm$0.05 & --- & --- & --- & +0.00$\pm$0.02 & +0.01$\pm$0.08\\ 
$[$Si/Fe$]$ & +0.09$\pm$0.03 & --- &+0.07$\pm$0.12  & +0.09$\pm$0.09 &+0.09$\pm$0.03 & +0.21$\pm$0.03\\ 
$[$TiI/Fe$]$ & --0.12$\pm$0.02 & --- & ---  & --0.05$\pm$0.10 & --0.01$\pm$0.01 & --0.14$\pm$0.03\\
$[$TiII/Fe$]$& --0.03$\pm$0.06 & --- & ---  & --0.04$\pm$0.15& --- & --- \\
$[$V/Fe$]$ & +0.02$\pm$0.04 & --- & --- & --- & +0.01$\pm$0.02 & ---\\
$[$Y/Fe$]$ & --0.12$\pm$0.05 & --- & --0.11 & --- & +0.01$\pm$0.01 & +0.10$\pm$0.00\\
\hline 
Star  &  \multicolumn{6}{c}{041 ($\delta$ tau)}  \\
\hline
T$_eff$ (K)& 4800 & 4938 & 4975 & --- & 4980 & 4875\\
log$g$ (dex)  & 2.8 & 2.69 & 2.65 & --- & 2.83  & 2.4\\
$v_t$ (km~s$^{-1}$)& 1.4 & 1.40 & 1.4 & --- & 1.25 & 2.0  \\
\hline        
$[$FeI/H$]$ & +0.10$\pm$0.01 & +0.14$\pm$0.07 & +0.11 & --- & +0.19$\pm$0.01 & +0.07$\pm$0.01\\
$[$FeII/H$]$& +0.13$\pm$0.03 & +0.26$\pm$0.16 & +0.07 & ---- & +0.18$\pm$0.03 & +0.04$\pm$0.02\\
\hline
$[$Al/Fe$]$ & +0.00$\pm$0.05 & +0.16$\pm$0.01 & --- & --- & +0.08$\pm$0.01 & --0.14$\pm$0.01\\
$[$Ba/Fe$]$ & +0.39$\pm$0.05 & --- & --0.02 & ---- & +0.15$\pm$0.01 & --0.13$\pm$0.00 \\
$[$Ca/Fe$]$ & --0.06$\pm$0.03 & --- &+0.08$\pm$0.12& --- & +0.00$\pm$0.05 & --0.17$\pm$0.06\\
$[$Co/Fe$]$ & +0.01$\pm$0.03 & --- & ---& --- & +0.01$\pm$0.03 & +0.10$\pm$0.05\\
$[$Cr/Fe$]$ & +0.03$\pm$0.03 & ---  & ---& --- & --0.04$\pm$0.02 & --0.06$\pm$0.09\\
$[$La/Fe$]$ & --0.08$\pm$0.05 & --- & --0.33 & --- & --0.05$\pm$0.09 & --- \\
$[$Mg/Fe$]$ & +0.06$\pm$0.04 & +0.36$\pm$0.02 & --0.10 & --- & +0.15$\pm$0.01 & +0.33$\pm$0.07 \\
$[$Na/Fe$]$ & +0.18$\pm$0.02 & +0.44$\pm$0.05 & +0.16 & --- & +0.32$\pm$0.01 & +0.28$\pm$0.05\\
$[$Nd/Fe$]$ & +0.08$\pm$0.30 & --- & --0.17 & --- & +0.02$\pm$0.02 & --0.17$\pm$0.00\\
$[$Ni/Fe$]$ & +0.04$\pm$0.02 &--0.03$\pm$0.06 &+0.06$\pm$0.09& --- & +0.09$\pm$0.06 & +0.11$\pm$0.03 \\
$[$O/Fe$]$  & --0.25$\pm$0.05 & --0.03$\pm$0.06 & --0.21 & --- & --- & --- \\
$[$Sc/Fe$]$ & +0.00$\pm$0.05 & --- & --- & --- & +0.01$\pm$0.02 & --0.02$\pm$0.07\\
$[$Si/Fe$]$ & +0.09$\pm$0.02 & --- & +0.07$\pm$0.11& --- & +0.06$\pm$0.02 & +0.23$\pm$0.03 \\
$[$TiI/Fe$]$ & --0.11$\pm$0.02 & --- & ---& --- & --0.04$\pm$0.02 & --0.07$\pm$0.03 \\
$[$TiII/Fe$]$& --0.00$\pm$0.07 & --- & --- & --- & --- & --- \\
$[$V/Fe$]$   & +0.00$\pm$0.03 & --- & ---& --- & +0.02$\pm$0.02 & ---  \\
$[$Y/Fe$]$  & --0.06$\pm$0.06 & --- & --0.10 & --- & --0.04$\pm$0.03 & +0.19$\pm$0.00 \\
\hline 
Star &  \multicolumn{6}{c}{070 ($\epsilon$ tau)} \\
\hline
T$_eff$ (K)   & 4800 & 4911 & 4925 & 4838 & 4880 & --- \\
log$g$ (dex) & 2.8 & 2.57 & 2.55 & 2.52 & 2.50 & ---  \\
$v_t$ (km~s$^{-1}$) & 1.6 & 1.47 & 1.4 & 1.63 & 1.46 & --- \\
\hline        
$[$FeI/H$]$ & +0.11$\pm$0.01 & +0.20$\pm$0.08 & +0.11  & +0.21$\pm$0.07 & +0.11$\pm$0.01 & ---  \\
$[$FeII/H$]$& +0.09$\pm$0.03 & +0.22$\pm$0.16 & +0.11  & +0.18$\pm$0.10 & +0.05$\pm$0.03 & ---  \\
\hline
$[$Al/Fe$]$ & +0.02$\pm$0.05 & +0.15$\pm$0.01 & ---  & +0.17$\pm$0.08 & +0.20$\pm$0.01 & --- \\
$[$Ba/Fe$]$ & +0.31$\pm$0.05 & --- & --0.02 & --- & +0.09$\pm$0.01 & --- \\
$[$Ca/Fe$]$ & --0.07$\pm$0.02 & --- & +0.11$\pm$0.12  & +0.01$\pm$0.10 & +0.09$\pm$0.03 & --- \\
$[$Co/Fe$]$ & +0.06$\pm$0.03 & --- & --- & --- & --0.01$\pm$0.04 & --- \\
$[$Cr/Fe$]$ & +0.08$\pm$0.04 & ---  & --- & --- & +0.00$\pm$0.01  & --- \\
$[$La/Fe$]$ & --0.05$\pm$0.05 & --- & -0.20 & --- & -0.17$\pm$0.00  & ---\\
$[$Mg/Fe$]$ & +0.21$\pm$0.07 & +0.37$\pm$0.02 & --0.08  & --0.03$\pm$0.08 & --- & --- \\
$[$Na/Fe$]$ & +0.18$\pm$0.02 & +0.41$\pm$0.04 & +0.23  & +0.04$\pm$0.11 & +0.40$\pm$0.04 & --- \\
$[$Nd/Fe$]$ & +0.08$\pm$0.28 & --- & --0.21 & --- & --0.10$\pm$0.05 & --- \\
$[$Ni/Fe$]$ & +0.03$\pm$0.02 & +0.06$\pm$0.08 & +0.09$\pm$0.11 & --- & +0.00$\pm$0.02 & --- \\
$[$O/Fe$]$ & --0.22$\pm$0.07 & --0.13$\pm$0.06 & --0.01  & --0.04$\pm$0.13 & --- & --- \\
$[$Sc/Fe$]$  & --0.04$\pm$0.05 & --- & --- & --- & ---  & --- \\
$[$Si/Fe$]$ & +0.10$\pm$0.03 & --- & +0.09$\pm$0.11  & +0.05$\pm$0.11 & +0.09$\pm$0.03 & --- \\
$[$TiI/Fe$]$& --0.06$\pm$0.01 & --- & ---  & --0.01$\pm$0.08 & --0.05$\pm$0.03  & ---\\
$[$TiII/Fe$]$& --0.02$\pm$0.11 & --- & ---  & --0.14$\pm$0.11 & ---  & ---\\
$[$V/Fe$]$  & +0.09$\pm$0.03 & --- & --- & --- & --0.04$\pm$0.03  & ---\\
$[$Y/Fe$]$  & --0.07$\pm$0.05 & --- & --0.11 & --- & --0.05$\pm$0.03  & ---\\
\hline 
\end{tabular}
\end{minipage}
\end{table*}

To validate the whole procedure used here, in Paper~I we performed an
abundance analysis of the ESO HARPS solar spectrum reflected by
Ganymede.
We used the same line list, model atmospheres, and abundance calculation code
that we used on our OC target stars, and found solar values for all elements,
with the only marginal exceptions of barium and aluminium (see also Section
\ref{sec6}). While the details of this analysis can be found in Paper~I, we
mention here that our reference solar abundances are taken from \citet{gre96}.

\section{Cluster-by-cluster discussion}\label{sec5}

\subsection{Berkeley 32}

Berkeley 32 ($\alpha_{2000}=06^h58^m07^s$ and $\delta_{2000}=+06^o25'43``$) is a
distant OC (R$_{gc}$=11.6 kpc) located towards the Galactic anticentre and
situated 260 pc above the disc plane. Its distance makes it one of the crucial
clusters for a correct determination of the metallicity gradient along the
Galactic disc, and therefore one of the key OC to the understanding of disc
formation and evolution. The color-magnitude diagram of this cluster
\citep[e.g.][]{dorazi2006}, contaminated by disc stars, shows a clear main
sequence turn-off with a sparsely populated red giant branch. Determinations of
its age, mainly using morphological indicators, yield a value of $\simeq$5 Gyr
\citep[e.g.][]{dorazi2006,salaris2004,Richtler2001,carraro1994,kaluzny1991}.

Given its large distance, it has not been well-studied spectroscopically, but we
could compare our results with two recent high-resolution studies of \citet{bragaglia2008} and \citet{friel2010}. We found a very close
agreement of our abundance ratios with those studies (see Table~\ref{be32}). The exceptions are Ba and Na. It is well-known that Ba abundances are enhanced by HFS \citep[e.g.][]{dorazi2009} effects that should explain the differences from \citet{bragaglia2008}. The [Na/Fe] ratio is lower than the values reported by \citet{bragaglia2008} and \citet{friel2010} by -0.25 and -0.32 dex, respectively. The difficulty in measuring Na lines, which suffer from NLTE effects, could easily explain this controversy. Moreover, different model atmospheres, stellar and atomic parameters, etc., between different studies may also play a role (and remove this discrepancy).

\subsection{NGC~752}

NGC~752 ($\alpha_{2000}=01^h57^m41^s$, $\delta_{2000}=+37^o47'06``$) is an old
($\sim$1.6 Gyr) OC located in the solar neighbourhood at a distance of $\simeq$400
pc. This cluster has a low central concentration and contains a relatively small
number of members. Its color-magnitude diagram \citep[e.g.][]{johnson1953} has a
still poorly understood morphology. The turn-off area is well-populated by early
F-type stars, while the low main-sequence appears to be sparsely populated (Figure
\ref{CMDs}). This, together with the age of this cluster, may be an
indication of the dynamic escape of low mass stars.  Stellar evolution models also
predict a well-populated red giant branch, which is not observed. All the known
red giants are located in the red clump region \citep{bartasiute2007}, which has a
peculiar morphology because it has a faint extension slightly to the blue of
its main concentration, which cannot be reproduced by stellar evolution models
\citep{girardi2000}.

\begin{table}
\begin{minipage}[htb]{\columnwidth}
\caption{High-resolution average Hyades abundances from giants.}
\label{hyades_cl}
\centering
\renewcommand{\footnoterule}{}
\begin{tabular}{l r r r r}
\hline\hline 
        & {\em Here} & 
        S09/S06\footnote{\citet{schuler2009} and \citet{schuler2006}, from the same 3 K giants studied here.} & 
	Bo00\footnote{\citet{boyarchuk2000}, from the same 3 K giants studied here.} &
	V99\footnote{\citet{varenne1999}, from 29 F dwarfs.}  \\
\hline
R=$\lambda/\delta\lambda$ & 30000   & 60000   & 45000 & 30--65000  \\
S/N                       & 50--100 & 100--200 & 100--300 & $\sim$200 \\
\hline
$[$Fe/H$]$ & +0.11$\pm$0.01 & +0.21$\pm$0.04 & +0.12$\pm$0.06 & --0.05$\pm$0.03 \\
$[$Al/Fe$]$ & +0.00$\pm$0.02 & +0.17$\pm$0.03  & +0.13$\pm$0.06 & ---  \\
$[$Ba/Fe$]$ & +0.36$\pm$0.04 &  --- & +0.10$\pm$0.03 & --0.03$\pm$0.07 \\
$[$Ca/Fe$]$ & --0.07$\pm$0.01 & --- & +0.03$\pm$0.05 & +0.03$\pm$0.04 \\
$[$Co/Fe$]$ & +0.03$\pm$0.03 & --- & +0.00$\pm$0.02 & --- \\
$[$Cr/Fe$]$ & +0.04$\pm$0.03 & --- & --0.02$\pm$0.02 & ---\\
$[$La/Fe$]$ & --0.08$\pm$0.04 &   ---  & --0.10$\pm$0.08  & ---\\
$[$Mg/Fe$]$ & +0.10$\pm$0.08 & +0.38$\pm$0.03 & +0.16$\pm$0.01 & +0.17$\pm$0.04\\
$[$Na/Fe$]$ & +0.18$\pm$0.01 & +0.45$\pm$0.04 & +0.37$\pm$0.05 & +0.10$\pm$0.05\\
$[$Ni/Fe$]$ & +0.03$\pm$0.01 & +0.05$\pm$0.07 & +0.05$\pm$0.06 & +0.04$\pm$0.04\\
$[$O/Fe$]$  & --0.27$\pm$0.04 & --0.08$\pm$0.05 & --- & +0.11$\pm$0.04\\
$[$Sc/Fe$]$ & --0.02$\pm$0.02 & --- &+0.00$\pm$0.01 & +0.03$\pm$0.07 \\
$[$Si/Fe$]$ & +0.09$\pm$0.01 & --- & +0.07$\pm$0.01 & +0.13$\pm$0.03\\
$[$Ti/Fe$]$ & --0.09$\pm$0.04 & --- &  --0.04$\pm$0.02 & ---\\
$[$V/Fe$]$ & +0.04$\pm$0.05 & --- & --0.01$\pm$0.03 & --- \\
$[$Y/Fe$]$ & --0.09$\pm$0.03 & --- & --0.04$\pm$0.02 & --0.04$\pm$0.07 \\
\hline
\end{tabular}
\end{minipage} 
\end{table}

Photometry and low/medium resolution spectroscopy studies \citep[see][and
references therein]{bartasiute2007} determined a slightly subsolar metallicity
\citep[i.e. {[}Fe/H{]}=--0.16$\pm$0.05,][]{friel1993}. A similar result was found
with high-resolution spectroscopy (R$\simeq$40\,000, S/N$\simeq$80--150) in eight
F-type stars around the main sequence turn-off
\citep[{[}Fe/H{]}=--0.09$\pm$0.05,][]{hobbs1992}. However, an investigation
based on high-resolution spectroscopy (R $\simeq$57\,000, S/N$\simeq$30--80) of 18
G giant stars obtained a solar [Fe/H] ratio
\citep[{[}Fe/H{]}=+0.01$\pm$0.04,][]{sestito2004} in closer agreement with the
value determined here. To our knowledge, we are the first to publish abundance
ratios of elements other than [Fe/H] for this cluster. 

\subsection{Hyades}

The Hyades cluster (Melotte 25; $\alpha_{2000}=04^h26^m54^s$ and
$\delta_{2000}=+15^o52'00``$) is the closest OC to the Sun ($\sim$45 pc) located
in the constellation of Taurus. Its proximity has motivated an extensive study
lasting more than a century \citep[starting with][]{herttzsprung1909}. The OC is
embedded into a moving group with the same name, which suggests that it would be
part of a dynamical stream coming from the inner Galaxy or a disrupting cluster
\citep{famaey2007}. 

Being one of the most studied clusters, both photometrically and
spectroscopically, it is the ideal cluster for abundance analysis comparisons.
The color-magnitude diagram of this young OC \citep[$\sim$0.7 Gyr, see Table \ref{phot}; e.g.][]{johnson1955}
contains only four red giant stars that have been confirmed as members from their
parallaxes, proper motions, and radial velocities. Most of the existing abundance
studies are focused on main sequence stars \citep[see e.g.][ and references
therein]{paulson2003,burkhart2000,varenne1999}. A comparison of the Hyades average
abundances determined from some (or all) of the known four red giants are shown
in Table~\ref{hyades_cl}. The averages of the abundances compiled until 1999 by
\citet{varenne1999} are shown in the last column of Table~\ref{hyades_cl} for
reference. In general, [Fe/H] appears slightly supersolar, while all other
abundance ratios are solar, and our abundance ratios agree well with literature
values. 

\begin{table}
\begin{minipage}[htb]{\columnwidth}
\caption{High-resolution average Praesepe (NGC~2632) abundances.}
\label{m44}
\centering
\renewcommand{\footnoterule}{}
\begin{tabular}{l r r r r r}
\hline\hline
        & {\em Here} & 
	P08\footnote{\citet{pace2008}, from 6 G and 1 F main sequence stars.} &
	A07\footnote{\citet{an2007}, from 4 G dwarfs stars.} &
        Bu98\footnote{\citet{burkhart1998}, from 10 Am stars.}\\
\hline
R=$\lambda/\delta\lambda$ & 30000 & 100000 & 55000  & 90000  \\
S/N                       & 50--100 & $\simeq$80 & $\simeq$100 & $\sim$200 \\
\hline
$[$Fe/H$]$  & +0.16$\pm$0.05 & +0.27$\pm$0.10 & +0.11$\pm$0.03 & +0.40$\pm$0.14 \\
$[$Al/Fe$]$ & +0.00$\pm$0.03 & --0.05$\pm$0.12 & --- & --0.19$\pm$0.17 \\
$[$Ba/Fe$]$ & +0.33$\pm$0.05 & +0.22$\pm$0.06 & --- & --- \\
$[$Ca/Fe$]$ & --0.08$\pm$0.02 & +0.00$\pm$0.11 & --- & +0.04$\pm$0.16 \\
$[$Co/Fe$]$ & +0.04$\pm$0.02 & --- & --- & --- \\
$[$Cr/Fe$]$ & +0.05$\pm$0.01 & --0.01$\pm$0.08 & --- & --- \\
$[$La/Fe$]$ & --0.05$\pm$0.02 & --- & --- & ---  \\
$[$Mg/Fe$]$ & +0.27$\pm$0.05 & --- & --- & --- \\
$[$Na/Fe$]$ & +0.25$\pm$0.06 & --0.04$\pm$0.12 & --- & ---\\
$[$Ni/Fe$]$ & +0.02$\pm$0.02 & --0.02$\pm$0.12 & --- &  +0.21$\pm$0.17 \\
$[$O/Fe$]$  & --0.11$\pm$0.03 & --0.40$\pm$0.20 & --- & --- \\
$[$Sc/Fe$]$ & --0.04$\pm$0.05 & --- & --- & --- \\
$[$Si/Fe$]$ & +0.06$\pm$0.02 & --0.01$\pm$0.12 & --- & --- \\
$[$Ti/Fe$]$ & --0.07$\pm$0.03 & --0.04$\pm$0.12 & --- & --- \\
$[$V/Fe$]$  & +0.06$\pm$0.03 & --- & --- & --- \\
$[$Y/Fe$]$  & --0.11$\pm$0.01 & --- & --- & --- \\
\hline
\end{tabular}
\end{minipage}
\end{table}

The three late-type Hyades giants (028, 041, and 070)  have been widely studied
\citep[e.g.][]{luck1995,boyarchuk2000,schuler2006,schuler2009,mishenina2006,mishenina2007,fulbright2007}.
In Table~\ref{tab_hyades_individual} we compiled available literature data. Our
temperatures are slightly lower (by $\sim$100 K) than the literature ones, whereas
our values of $\log g$ and $v_t$ are similar. These marginal differences appear to have no
significant impact on the derived abundance ratios, which agree very well with
literature ones. Exceptions are Al, Ba, and O, which suffer from technical measurement
problems (not strictly related to the Hyades cluster) and are discussed in
Sections~\ref{sec-err} and \ref{sec6}.

\subsection{Praesepe (NGC~2632)}

The cluster popularly known as Praesepe or Beehive (also called M~44, NGC~2632 or
Melotte 88; $\alpha_{2000}=08^h40^m24^s$ and $\delta_{2000}=+19^o40'00``$ ), is an
old OC (0.65 Gyr, see Table~\ref{phot}) well known from the antiquity. It is
located in the Cancer constellation at a distance of $\simeq$175~pc, computed from
Hipparcos parallaxes.

Its metal content was derived with different methods
\citep[e.g.][]{friel1992,komarov1993,claria1996,huibonhoa1998,burkhart1998,burkhart2000,dias2002,pace2008}.
In general, all the quoted studies obtained a metallicity either barely or
definitely supersolar. Of these, the high-resolution abundance determinations
were derived mainly for dwarfs or early-type giants
\citep[e.g.][]{friel1992,burkhart1998,an2007,pace2008}. Surprisingly, to our
knowledge, there are no recent high-resolution abundance determinations of
late-type giants in this cluster. 

Table \ref{m44} shows a comparison of our results with some of the most recent
high-resolution studies. In general, the [Fe/H] we derived in our late-type giants
lies in-between those of \citet{pace2008} and \citet{an2007}, suggesting that the
proposed dichotomy of literature values (barely supersolar versus definitely
supersolar) should be interpreted rather as an above average uncertainty. This larger
than usual uncertainty could naturally arise from the different spectral types and
abundance analysis methods employed in the literature. The [Fe/H] ratio derived by
\citep{burkhart1998}, based on Am stars, is on average $\simeq$0.3 dex larger than the
values obtained in other works using different spectral-type stars. Although these
stars should in principle reflect the chemical composition of the cluster, Am stars
always have overabundant Fe abundances relative to other objects in the same
clusters, without a clear explanation appearing in the literature. As in the case of
the Hyades, Na and O abundances derived by us appear marginally discrepant with those
by \citet{pace2008}, and will be discussed in more detail in Section~\ref{sec6}.


\section{Discussion of abundance ratios}\label{sec6}

\begin{figure}
\centering
\includegraphics[width=\columnwidth]{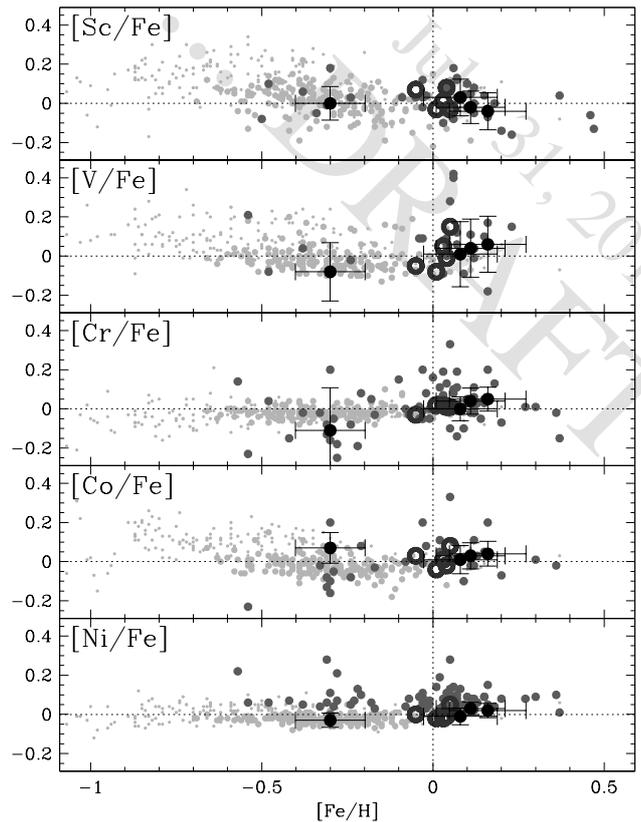}

\caption{Comparison between our iron-peak abundance ratios (large black dots),
those of Paper I (large black open circles), high-resolution measurements listed
in Table~\ref{tab-hiresnew} (large dark grey dots), field stars belonging to the
thin disc \citep[light grey dots,][]{reddy2003}, and to the thick disc
\citep[tiny light grey dots,][]{reddy2006}. Errorbars in our measurements are the
quadratic sum of all uncertainties discussed in Sections~\ref{sec4} and
\ref{sec-err}.}

\label{fig_fe_peak}
\end{figure}

As in Paper I, we compared our abundance ratios (and those from Paper~I) with
both others in the literature and the abundances of the Galactic disc field stars
from \citet{reddy2006,reddy2003} in Figures \ref{fig_fe_peak} to \ref{fig_naal}. We
extended the open cluster abundance compilation of Paper~I (see
Table~\ref{tab-hiresnew}) with both recent published works and old studies that were not included in the previous version. In both cases, as in Paper~I, we included only studies based on high-resolution (R$\gtrsim$18000) spectroscopy. When more than one determination was available
for one cluster, we simply plotted them all to give a realistic idea of the
uncertainties involved in the compilation, and we did not attempt to correct for
differences between the abundance analysis procedures (log$gf$, solar reference, and so
on), because this would be beyond the scope of the present article.

\subsection{Iron-peak element ratios}\label{iron-abo}

Figure~\ref{fig_fe_peak} shows the abundance ratios of iron-peak elements. Our OC
with abundances close to solar (i.e., Hyades, Praesepe, and NGC~752) are in very
good agreement with the results obtained in other OC studied with high-resolution
spectra and in disc stars of similar metallicity. A larger scatter or marginal
discrepancies are sometimes observed for the odd elements Sc, V, and Co, but this
is because of the well-known hyperfine structure (HFS) of the lines usually
employed in the analysis. The element that appears to suffer more from these
effects is vanadium. This scatter is due, at least in part, to the different procedures used in the literature for treating the HFS splitting. We stress that in our case, we do not attempt any HFS correction.

The most metal-poor and oldest OC in our sample, Be~32, has a puzzling
behaviour. While all its iron-peak abundance ratios are still compatible with the
literature values for OC and field stars of similar metallicity (uncertainties
are large), some underlying discrepancy could be present. For example, HFS should
cause an overestimate (and not an underestimate) of vanadium. In addition, chromium
appears to be lower than solar. We note that (see Table~\ref{be32})
the literature Co and Cr determinations by \citet{friel2010}, \citet{sestito2006},
and \citet{bragaglia2008} are very similar to ours. In the case of our [Cr/Fe] measurement for Be~32, we must note that our two giants appear to exhibit quite different [Cr/Fe] abundances, resulting in a large scatter in the cluster average value. This large scatter is most probably due to a measurement uncertainty, and should not be considered significant.

In Paper~I, we noticed a peculiar behaviour in the Ni abundance ratios of
literature OC determinations: they appear to be slightly richer in Ni than field
stars by roughly 0.05~dex. Our [Ni/Fe] ratios are in closer agreement with the
field star determinations than with the OC ones. Although this difference is
small (within the uncertainties), it appears systematic in nature, and we were 
unable to find any easy explanation, such as the choice of either solar reference
abundances or the log$gf$ system, of this discrepancy. 

\subsection{Alpha-element ratios}
\label{alfa-abo}

\begin{figure}
\centering
\includegraphics[width=\columnwidth]{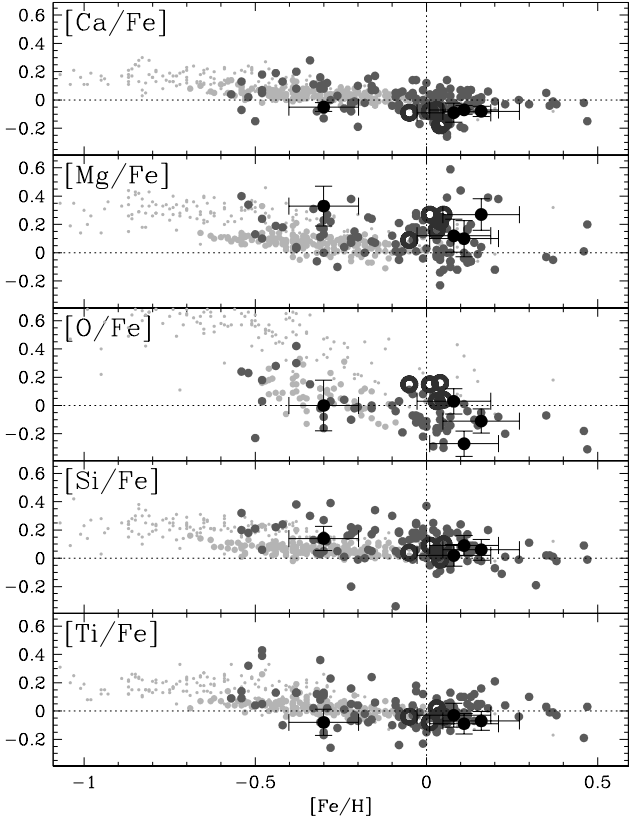}
\caption{Comparison of our $\alpha$-elements ratios with literature
values. Symbols are the same as in Figure~\ref{fig_fe_peak}.}
\label{fig_alfas}
\end{figure}

Figure~\ref{fig_alfas} shows the abundance ratios of $\alpha$-elements. As for iron-peak elements, our measurements are always compatible with the
literature values, within their uncertainties. Generally speaking, all our OC
show roughly solar $\alpha$-enhancements, even Be~32, which has a lower
metallicity. 

However, some elements deserve some more discussion, as was noted in Paper~I. For
example, the log$gf$ of calcium are quite uncertain, and we chose the VALD
reference atomic data, which explains why our [Ca/Fe] ratios are slightly lower
than the bulk of literature determinations for cluster and disc stars. A similar
problem affects the Mg lines, as can clearly be appreciated from the large spread
of literature values. Our [Mg/Fe] determinations tend to lie on the upper
envelope of literature ratios for OC. A deeper discussion of Mg abundances can be
found in Paper~I.

In the case of oxygen, the problem is instead in the difficulty in measuring its
small lines. The forbidden [O~I] line at 6300~\AA, which we used in this paper,
suffers from contamination by a Ni line and by telluric absorption features, while
the O triplet around 7770~\AA\  (used by some other studies) suffers from NLTE
effects. This is reflected by the large scatter in literature values.

\begin{figure}
\centering
\includegraphics[width=\columnwidth]{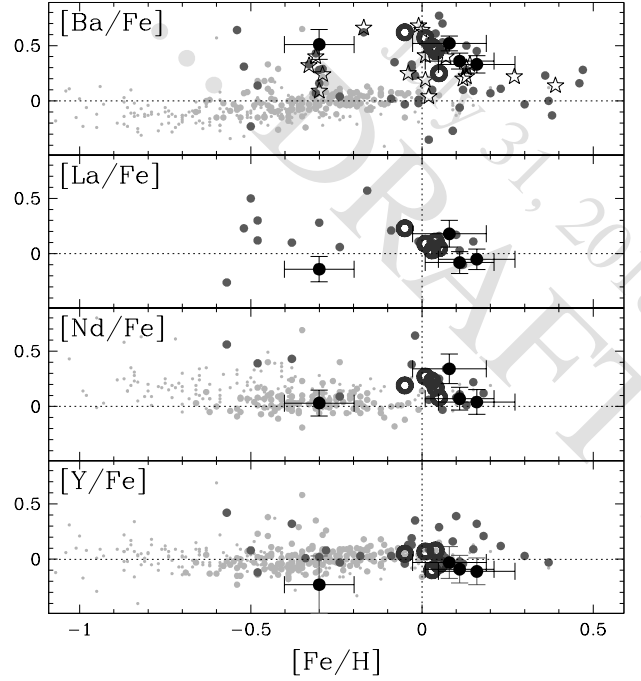}

\caption{Comparison of our s-process elements ratios with the literature
ones. Symbols are the same as in Figure~\ref{fig_fe_peak}, except for the black
star-like symbols in the top [Ba/Fe] panel, which represent the revision of Ba
abundances with spectral synthesis performed by \citet{dorazi2009}.}
\label{fig_esse}
\end{figure}

\subsection{Heavy element ratios}
\label{heavy-abo}

We determined abundances for three heavy s-process elements: Ba, La, and Nd; and
one light s-process element: Y (Figure~\ref{fig_esse}). Literature determinations
for these elements are not numerous. \citet{dorazi2009} measured Ba in several OC
using spectral synthesis to take into account HFS. The [Ba/Fe] abundances derived by \citet{dorazi2009} taking into account HFS do not differ significantly from other literature determinations (including ours). The [Ba/Fe] ratios are clearly above solar for most of the clusters and they show a scatter larger than $\sim$0.5. \citet{dorazi2009} found this scatter to be due to age: the Ba content appears to increase for younger clusters. The available lanthanum and neodymium lines were unfortunately relatively
small, and we were able to find fewer published studies to compare with. As a result, the
solar clusters (Hyades, Praesepe, and NGC~752) have La and Nd ratios in good
agreement with the literature, while Be~32 appears to have lower [La/Fe] and
[Nd/Fe] than the few studied OC at a similar metallicity, which are Mel~66
\citep{gratton1994} and NGC~2243 \citep{smith1987}. However, our [Nd/Fe] agrees
well with the field star solar ratios. The only light s-process element we could
measure, Y, relies on a couple of weak lines that provide uncertain abundances
(see the large errorbar in Figure~\ref{fig_esse}). Our Y ratio appears to be lower than
all literature estimates, although still compatible with the solar values of
field stars of similar metallicity, within the large uncertainties. 

In summary, we can say that all the studied clusters appear to have roughly solar
s-process enhancements, but it would be extremely interesting to attempt a more detailed study of s-process elements in OC, as done by
\citet{dorazi2009} for barium.

\subsection{Ratios of Na and Al and anticorrelations}

\begin{figure}
\centering
\includegraphics[width=\columnwidth]{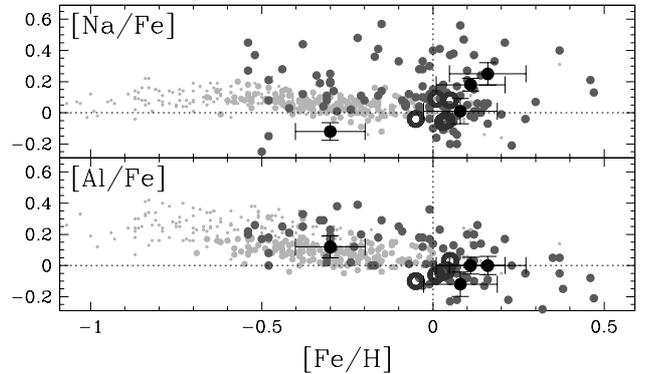}
\caption{Comparison between our [Na/Fe] and [Al/Fe] ratios and the literature
ones. Symbols are the same as in Figure~\ref{fig_fe_peak}.}
\label{fig_naal}
\end{figure}

As discussed in Paper~I, the study of light elements in OC is quite
interesting.  The elements Al and Na, together with Mg, O, C, and N, show puzzling
(anti-)correlations in almost all of the studied globular clusters, in the Milky
Way \citep[see, e.g.,][and references therein]{carretta2010,pancino2010b} and
outside \citep[e.g.][]{mucciarelli2009,letarte2006}. No (anti-)correlations were
observed in either field stars \citep[but see][]{martell2010} or OC \citep[][and
Paper~I]{martell2009,desilva2009,smiljanic2009} so far. This suggests that
metallicity, cluster size and age, or the environment must play a r\^ole, and
therefore finding (anti-)correlations in some OC would be of enormous
importance to put tighter constraints on the phenomenon.

We determined abundances of Al and Na and compared them with published
results in Figure~\ref{fig_naal}. While in the case of aluminium the agreement
with literature values is good, we find a significantly lower [Na/Fe] ratio for
Be~32 than for other clusters or field stars of similar metallicity. Generally
speaking, the large scatter in Na determinations could be due to the difficulties
in measuring Na lines, often affected by NLTE effects \citep{gratton1999},
although no such scatter is observed among field stars. However, a few clusters
have [Na/Fe] lower than our Be~32 determination, and NLTE corrections
\citep{gratton1999} could make the discrepancy of our Be~32 Na determination even
worse. Unfortunately, given the large scatter and the difficulty of measurement, it is difficult to either confirm or exclude the presence of some (small) intrinsic [Na/Fe] scatter in this clusters.

\begin{figure}
\centering
\includegraphics[width=\columnwidth,height=\columnwidth]{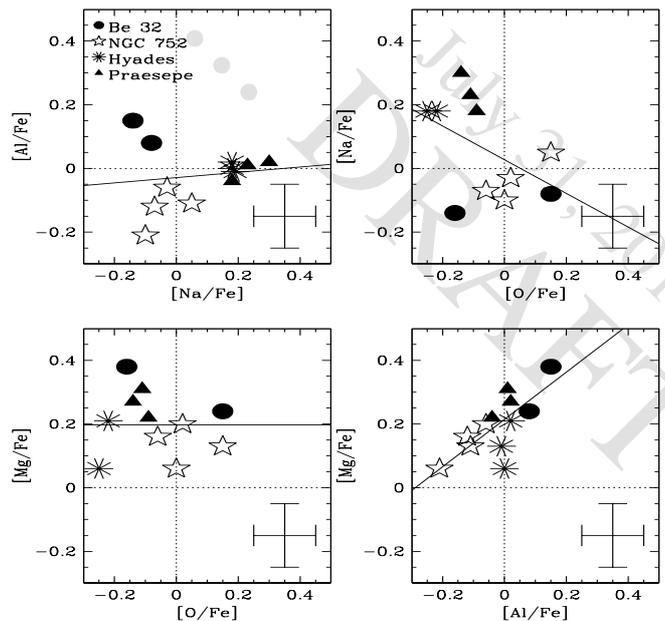}
\caption{A search for (anti)-correlations of Al, Mg, Na, and O among our target
stars. The four panels show different planes of abundance ratios, where stars
belonging to each cluster are marked with different symbols. Dotted lines show
solar values, solid lines show linear regressions and the typical uncertainty
($\sim$0.1~dex) is marked at the lower right corner of each panel.}
\label{fig_anticorr}
\end{figure}

In Figure~\ref{fig_anticorr}, all the studied stars occupy the ``normal stars"
loci, which is around solar for Na and Al, and slightly $\alpha$-enhanced for O
and Mg (see Section \ref{alfa-abo}). There is a hint of correlation between
[Al/Fe] and [Na/Fe], which was also observed for objects studied in Paper I. Of
course, small variations in T$_{\rm{eff}}$ could induce artificial correlations
between element pairs, so the observed trend is most probably not-significant.
However, we again note that the Na spread is very large (see above), suggesting
that a small degree of chemical anomalies (barely hidden within the present
observational uncertainties) cannot be completely excluded.

\section{Galactic trends}\label{sec7}

The existence of trends in the chemical abundances with Galactocentric distance,
$R_{gc}$, vertical distance to the Galactic plane, $z$, and age, are key to
understanding Galactic disc formation and evolution because they provide
fundamental constraints on chemical evolution models. Different tracers have been used
to investigate trends in the Galactic disc: OB stars \citep[e.g.][]{daflon2004},
Cepheids \citep[e.g.][]{lemasle2008}, H II regions \citep[e.g.][]{deharveng2000},
and planetary nebulae \citep[e.g.][]{costa2004}. However, as coeval groups of
stars at the same distance and with a homogeneous chemical composition, OC are
the ideal test particles to investigate the existence of radial and vertical
gradients and of an age-metallicity relation in the disc.

We complement the small sample of abundance ratios obtained here and in Paper~I
with a revised version of the literature data first presented in Paper~I
(Table~\ref{tab-hiresnew}). When a cluster had two or more abundance
determinations available in the literature, we averaged them to make the figures
easily readable and the error bars are, simply, calculated as the standard deviation. For those clusters with only one abundance determination, the error bars are the uncertainties in those determinations. The heliocentric distances
compiled in the updated version of the \citet{dias2002} database were used to
obtain $R_{gc}$ and $z$ for each cluster, assuming $R_{GC_{\odot}}$=8.5~kpc. Ages
were obtained from the same source, which is a compilation of different values
available in the literature, hence might still be quite inaccurate for
some clusters. In spite of its heterogeneity, our compilation contains a total of 89 clusters and is, to the best of our
knowledge, the largest available in the literature, based on high-resolution
spectroscopic abundances. Any attempt to homogenize this sample, for which abundances, distances, and ages have been derived from very different techniques, is clearly beyond the scope of this paper. This prevents us from a detailed analysis of the Galactic trends of all elements. For this reason, we focus only on [Fe/H] and [$\alpha$/Fe] ratios. In spite of this heterogeneity, this analysis is still very useful owing to the number of clusters, and the large range of ages, and vertical and radial distances covered, even if the heterogeneity of the sample forces us to be extremely cautions when drawing any conclusion from the data.

\begin{figure}
\centering
\includegraphics[width=\columnwidth,height=\columnwidth]{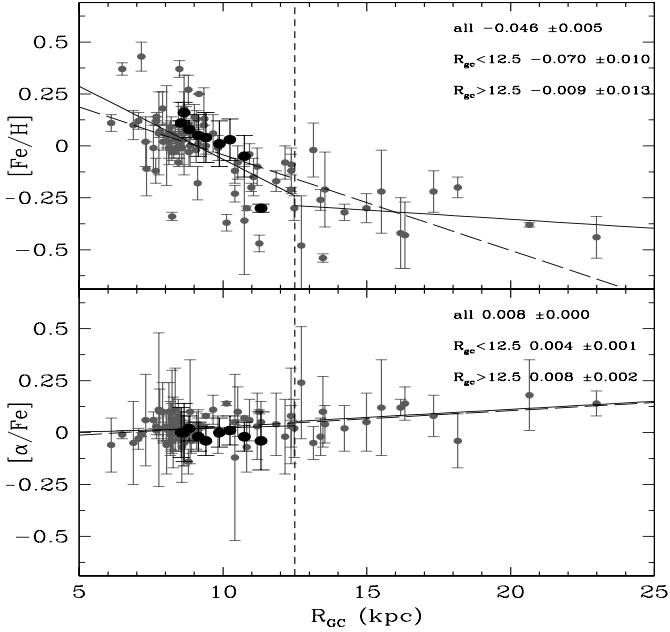}
\caption{Trends of [Fe/H] (top panel) and [$\alpha$/Fe] (bottom panel) with
galactocentric radius. Grey dots are OC compiled in Table~\ref{tab-hiresnew},
while black dots are the ones analysed here and in Paper I. A global linear fit
is drawn in both panels (long-dashed line). Two separate linear fits of OC inside
and outside 12.5~kpc are also shown (solid lines).}
\label{fig_trend_rgc}
\end{figure}

\subsection{Trends with Galactocentric radius}

Radial gradients may arise when the disc forms, and different mechanisms can
produce them: for example, different timescales of star formation at different
distances \citep[e.g.,][]{schaye2004}; a radial variation in the infall of gas; or
a change in the yield as a function of the radius \citep[e.g.,][]{molla1996}.
This initial radial gradient can be either amplified (steepened) or washed out
(flattened) with time by radial mixing \citep[e.g.,][]{roskar2008}.

Since the pioneering work of \citet{janes1979}, OC have been widely used to
investigate the gradient in metallicity with radius in the Galactic disc
\citep[e.g.,][]{twarog2003,friel02,magrini2009,friel2010,jacobson2011,jacobson2011b}.
\citet{friel1995} reviewed the firsts investigations in this field. Since then, a
great effort have been performed to obtain both homogeneous
\citep[e.g.,][]{friel02,sestito2008,friel2010} and/or larger samples   
\citep[e.g.,][]{twarog1997,jacobson2011,jacobson2011b}. All these investigations agree on the
fact that the iron content decreases with increasing radius
\citep[e.g.,][]{friel02}. This behaviour has been generally considered linear
with a slope between --0.05 and --0.09 dex kpc$^{-1}$, depending on the cluster
sample used. Similar trends were obtained for other different tracers of the disc
\citep[e.g.,][]{andrievsky2004,lemasle2008}. Most of these works were limited to
the inner $R_{gc}\simeq$15 kpc. However, investigations based on samples
containing clusters at larger distances
\citep[e.g.,][]{twarog1997,yong2005,sestito2008} found that the [Fe/H] ratio
decreases as a function of increasing radius to $R_{gc}\simeq$12.5 kpc and
appears to flatten from there outwards.

The variation in [Fe/H] with $R_{gc}$ in our compilation has been plotted in the top
panel of Figure~\ref{fig_trend_rgc}. The whole sample is well fitted by a line
with a slope of --0.046$\pm$0.005 dex kpc$^{-1}$ (long-dashed line), in
concordance with the result obtained in Paper I from a $\simeq$20\% smaller
sample (--0.05$\pm$0.01 dex kpc$^{-1}$) and in other investigations in the
literature \cite[e.g. --0.06$\pm$0.02 dex kpc$^{-1}$;][]{friel02}. The sample
used here contains more clusters with distances larger than $R_{gc}\geq$12 kpc.
This allows us to investigate the discontinuity observed by some authors at
$R_{gc}\simeq$12-13 kpc. At first sight, no clear discontinuity in slope appears,
partly because of the large range of [Fe/H] at this radius ($\simeq$0.5 dex) and
partly as a possible consequence of the heterogeneity of our sample. However,
when we fit separately clusters inwards and outwards of 12.5 kpc, we find two
significantly different slopes: the metallicity in the inner disc decreases with
a slope of --0.07$\pm$0.01 dex kpc$^{-1}$, while in the outer disc the slope is
--0.01$\pm$0.01 dex kpc$^{-1}$. The obtained slopes change within the uncertainties if the cut radius varies between 11.5 and 13.5 kpc. This is also in very good agreement with the
recent results by \citet{andreuzzi2011}, who find --0.07 dex kpc$^{-1}$ in the
inner 12 kpc. This bimodal behaviour can be explained by a different chemical
enrichment and star formation in the inner and outer disc;
\citep[e.g.][]{chiappini2001,magrini2009} however, a sharp discontinuity between
the inner and outer disc is not expected theoretically. 

The ratio [$\alpha$/Fe] reflects the relative contributions of Type Ia and II
supernovae: chemical evolution models predict an increase of this ratio with
$R_{gc}$ \citep[e.g.][]{chiappini2001,magrini2009}. This tendency was indeed
observed in OC by, e.g., \citet{yong2005}, \citet{magrini2009}, and in Paper~I.
The bottom panel of Figure~\ref{fig_trend_rgc} shows the variation in [$\alpha$/Fe]
with $R_{gc}$ for our compilation: a weak increase in $\alpha$-element abundances
with radius is apparent. However, the slope is still compatible with a flat
distribution at the 1\,$\sigma$ level, as in Paper~I, especially if the two outermost clusters are removed. The discontinuity observed
for [Fe/H] is not evident at all in [$\alpha$/Fe].

An accretion of a satellite into the outer disc could also explain the trend
observed \citep[e.g.][]{chiappini2001,yong2005}. In this case, we would expect to find 
some inhomogeneities corresponding to the trajectory of the merger. \citet{carraro2009} indeed found evidence that two OC, Berkeley~29 and Saurer 1, are
related to the Sagittarius dwarf galaxy. Our compiled sample unfortunately do
not allow us to investigate this question in depth.

\begin{figure}
\centering
\includegraphics[width=\columnwidth,height=\columnwidth]{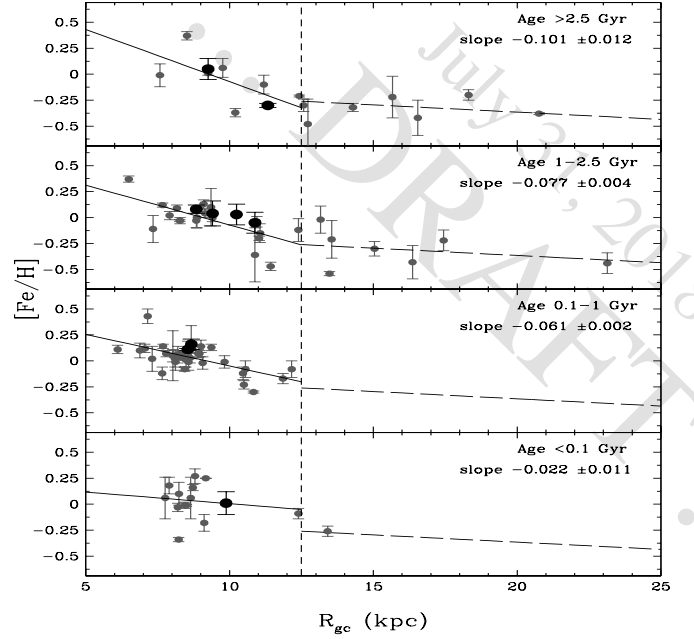}
\caption{Gradient in [Fe/H] as a function of $R_{gc}$ in four different age bins (labeled
in top--right corner). A linear fit is performed for the OC within a radius of
$R_{gc}$=12.5~kpc, and the slope indicated on each panel. A flatter and roughly
constant slope is found outside a radius of $R_{gc}$=12.5~kpc.}
\label{fig_trend_slope}
\end{figure}

\subsection{Time evolution of the radial gradient}

Chemical evolution models of the Galactic disc predict a variation in the metallicity
gradient with time, but they disagree about the direction of this gradient variation
\citep[see][for a recent review]{maciel2007}, some predicting a steepening and some a
flattening of the gradient with time. Studies based on metallicities derived from
low-resolution spectroscopy found that old OC ($\gtrsim$1 Gyr) followed a steeper
radial gradient, $\sim$-0.08 dex kpc$^{-1}$, than the younger ones, $\sim$-0.02 dex
kpc$^{-1}$ \citep{friel02,chen2003}. Only recently have chemical abundances been
derived from high-resolution spectroscopy for a sufficient number of OC to significantly
investigate the variation in the radial gradient with time. As for studies
based on low-resolution spectra, they agree that the gradient was steeper in the
past and has flattened with time
\citep{magrini2009,andreuzzi2011}. For example, on the basis of a sample of $\sim$70 OC
\citet{andreuzzi2011} found that all objects younger than 4 Gyr display a similar
gradient with a slope -0.07  dex kpc$^{-1}$ in the inner 12 kpc, while the one for
older objects is steeper, -0.15 dex kpc$^{-1}$. 

Other tracers have been used to study the time variation in radial gradients. Studies
based on planetary nebulae found more puzzling results: while \citet{maciel2003} found
a flattening of the gradient with time, as generally observed for OC,
\citet{stanghellini2010} found that the gradient steepens with time. At the moment,
there is no explanation of this contradictory result. Comparisons among the slopes of
the radial gradients described by populations of different ages also show that the
gradient has flattened out in the past few Gyr \citep[see][for a recent
review]{maciel2009}.

To investigate the behaviour of the radial gradient in our compiled sample of
high-resolution  abundances, we plotted in Figure~\ref{fig_trend_slope}  the gradient in [Fe/H] as a function of $R_{gc}$ in four different age bins. We obtained a
linear fit in each age bin for the inner 12.5 kpc, and for the outer range we
simply used the same fit as in Figure~\ref{fig_trend_rgc}, owing to the paucity of
OC after age binning in this region. We found that the slope of the [Fe/H]
gradient increases as we go back in time from --0.02$\pm$0.01 dex kpc$^{-1}$ for
objects younger than 0.1~Gyr to --0.10$\pm$0.01 dex kpc$^{-1}$ for clusters older
than 2.5~Gyr.

\begin{figure}
\centering
\includegraphics[width=\columnwidth,height=\columnwidth]{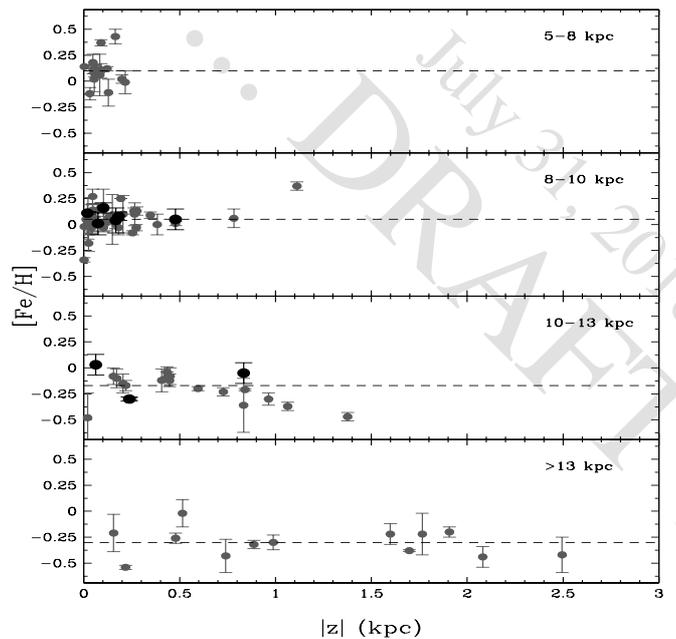}
\caption{Trends in [Fe/H] with $|z|$ in four radial annuli as indicated on the top-right
corner of each panel, moving outwards from the top to the bottom panel. Symbols are the
same as in Figure~\ref{fig_trend_rgc}. As a reference, we plotted dashed lines in
each panel, representing the median metallicity of clusters in each radial annulus.}
\label{fig_trend_z}
\end{figure}

\subsection{Trends with the disc scale--height}

Another interesting trend that could be investigated is the behaviour of [Fe/H] with
the vertical scale-height of the disc $z$, i.e., the vertical [Fe/H] gradient. 
Although the formation of the thick discs remains an open question, the existence of
vertical gradients can help us to discriminate among the mechanisms proposed to their
formation. No vertical chemical gradients are expected in thick discs formed by
heating caused by accretion events or major mergers. In contrast, vertical gradients
may exist in discs thickened by gradual heating of the thin disc or before the gas has
settled to form a thin disc \citep[see][for a review]{mould2005}. Up to now, there is
no conclusive agreement about the existence of a vertical metallicity gradient in
the Galactic disk. The existence of a vertical gradient for field stars have been
claimed by several authors, although they cover only about 1 kpc above and below the
disc plane \citep{bartasiute2003,marsakov2005,marsakov2006,soubiran2008}. Studies
covering large ranges of $|z|$ do not find any evidence of a vertical gradient
\citep{gilmore1995,soubiran2005,navarro2011} among the field populations. Studies
using OCs have found a vertical gradient of $\sim$-0.3dex
kpc$^{-1}$\citep{piatti1995,carraro1998,chen2003}, although, these studies do not
distinguish the effects of the radial gradient, which can mask any vertical trend. This
effects were taken into account by \citet{jacobson2011} who found no
evidence of a vertical gradient.

To investigate the presence of trends with $z$ in our compilation, we firstly had
to remove the contribution of the radial metallicity gradient. We plotted in 
Figure~\ref{fig_trend_z} the variation in [Fe/H] with $|z|$ in four different annuli of
$R_{gc}$. We note that OC with high $|z|$ are preferentially located at
large $R_{gc}$; this is not unexpected because the disc thickens in its external
regions. Moreover, an intrinsic bias caused by obscuration in the plane appears: clusters at large Galactocentric radii are found and observed preferentially higher above the plane. This could explain why the two outermost annuli studied uncover a possible weak decrease in [Fe/H] as $z$ increases.
This trend is however still compatible with no gradient at the 1\,$\sigma$ level
and, once again, larger samples of homogeneous data are necessary to investigate
this result in detail.  

\begin{figure}
\centering
\includegraphics[width=\columnwidth,height=\columnwidth]{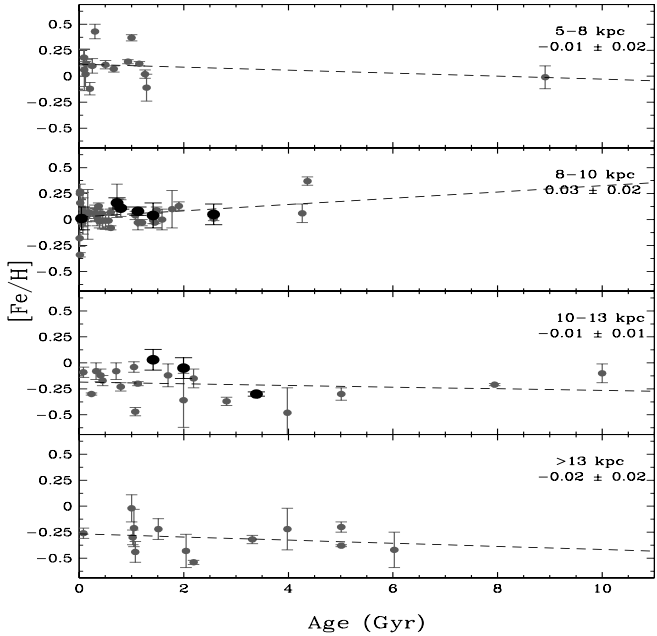}
\caption{The evolution of [Fe/H] with age in the same four radial annuli as in
Figure~\ref{fig_trend_z}. Again, dashed lines representing the median metallicity of clusters in each radial annulus have been plotted as reference.}
\label{fig_trend_age}
\end{figure}

\subsection{Is there an age--metallicity relation for open clusters?}

Another important prediction of the chemical evolution models is the existence of
an age-metallicity relation for disc populations. It is still unclear whether or not the field disc stars follow an age-metallicity relation. Some works find it \citep[e.g.][]{reddy2003,bensby2004,reid2007}, but others do not \citep[e.g.][]{feltzing2001,nordstrom2004,karatas2005}. Again, no clear trend of
chemical abundances with age has been clearly observed in the case of Galactic OC \citep[e.g.][]{friel1995}. Although \citet{friel2010} notices a trend of [Al/Fe]
and [O/Fe] ratios with age, again larger and homogeneous samples are necessary to
confirm this result. If an age-metallicity relation is confirmed for the field population but not for OC, this would imply that they might have followed a different chemical evolution \citep{yong2005}.

The evolution of the radial gradient as a function of time, described above,
indicates that the chemical enrichment of OC is modulated by their location in
the Galaxy and not by the moment at which they formed. To investigate whether an
age-metallicity relationship exits at a given $R_{gc}$, we plotted in
Figure~\ref{fig_trend_age} the evolution of [Fe/H] with age in four different radial
annuli. There is no clear trend in any of the studied annuli, although not all of
them contain clusters covering the same age range. Only in the outermost annulus
is a weak trend observed, although it is still not very significant. Again, we
conclude that a larger sample of homogeneous data are necessary to investigate this
point in depth.

\section{Summary and conclusions}\label{sec8}

We have enlarged our sample of homogeneous high-resolution abundance measurements
from the five clusters of Paper~I to a total of nine, analysing here spectra of
red clump giants in the Hyades, Praesepe, NGC~752, and Be~32. Our main results can be summarized as follows:

\begin{itemize}
\item{We provide the first high-resolution based abundance ratios \citep[other
than {[}Fe/H{]}, see][]{sestito2008} for NGC~752, which turned out to be mostly
of solar composition;} 
\item{We have presented the abundance ratios of Praesepe red clump giants, which appear to
solve a puzzling dichotomy of literature determinations for stars of different
evolutionary stages;}
\item{We have found that our abundance ratios for the Hyades and Berkeley~32 are in
good agreement with other literature determinations;}
\item{We have confirmed the absence of light elements (anti-)correlations in the OC
studied so far.}
\end{itemize}

We have updated our compilation of previous literature data for 57 clusters of Paper~I to
a total of 89 clusters presented here. With this updated compilation and our
homogeneous measurements in hand, we have investigated Galactic trends in [Fe/H] (and
[$\alpha$/Fe]) with age, Galactocentric radius, and height above the Galactic plane.
Our findings are in substantial agreement with other similar investigations, where the
abundance gradient appears to indeed flatten out outside $R_{gc}\simeq$12.5~kpc, and
the inner disc slope appears to flatten for younger ages as well, although the age
bins are not too well-sampled. At the same time, [$\alpha$/Fe] shows a weak increase
with $R_{gc}$. No significant gradients are observed with $|$z$|$ or age, except for a
weak tendency of [Fe/H] to decrease with increasing $|$z$|$ and decrease with age in
the outermost disc annulus studied. None of our measured weak trends have any significance
above 1\,$\sigma$. Larger samples of homogeneous data are still necessary to
investigate the existence of any dependence on age and $|$z$|$ in the Galactic disc.

\begin{acknowledgements}

We acknowledge the anonymous referee for helping us to improve this paper. R. C acknowledge the support from the Spanish Ministry of Science and Technology
(Plan Nacional de Investigaci\'on Cient\'{\i}fica, Desarrollo, e Investigaci\'on
Tecnol\'ogica, AYA2004-06343 and AYA2007-3E3507). R. C. also acknowledges the
funds by the Spanish Ministry of Science and Innovation under the Juan de la
Cierva and MEC/Fullbrigth fellowships, and by the Centro de Investigaciones de
Astronom\'{\i}a (Venezuela) under its postdoctoral fellowship programe.

\end{acknowledgements}

\onllongtab{12}{
\begin{landscape}
\begin{longtable}{lccccccl}
\caption{\label{tab-hiresnew}Literature sources of high-resolution (R$\geq$1500) [Fe/H] ratios of open clusters together with the resolution, signal-to-noise ratios, number of stars, and method used in each of them.}\\
\hline
\hline
Cluster & & [Fe/H] & Resolution & S/N & N. Star & Method &  References \\
\hline
\endfirsthead
\caption{Continued.} \\
\hline
Cluster & & [Fe/H] & Resolution & S/N & N. Star & Method &  References \\
\hline
\endhead
\hline
\endfoot
\hline
\endlastfoot
Be~17 & & --0.10$\pm$0.09 & 25000 & $\geq$80  & 3 giant &  EW\footnote[1]{Spectral synthesis for O.} & \citet{friel2005}\\
Be~20 & & --0.48$\pm$0.08 & 28000 & $\geq$50  & 2 giant &  EW & \citet{yong2005} \\
         & & --0.30$\pm$0.02 & 45000 & $\geq$35  & 2 giant &  EW & \citet{sestito2008} \\ 
Be~21 & & --0.54$\pm$0.02 & 48000 & $\sim$20 &4 giants & EW & \citet{hill1999}\\
Be~22 & & --0.32$\pm$0.04 & 34000 & 20--25    & 2 K giant & EW & \citet{villanova2005}   \\
Be~25 & & --0.20$\pm$0.05 & 40000 & 25--40    & 4 giant &  EW & \citet{carraro2007b}   \\
Be~29 & & --0.44$\pm$0.02 & 34000 & $\sim$70  & 2 G giant & EW & \citet{carraro2004}  \\
         & & --0.52$\pm$0.03 & 28000 & $\geq$100 & 2 giant &  EW & \citet{yong2005}  \\
         & & --0.31$\pm$0.03 & 45000 & $\geq$25  & 6 giant & EW & \citet{sestito2008} \\
Be~31 & & --0.54$\pm$0.06 & 28000 & $\geq$60  & 1 giant & EW & \citet{yong2005}  \\
         & & --0.31$\pm$0.06 & 28000 & $\sim$100 & 2 giant & EW\footnotemark[1]  & \citet{friel2010} \\
Be~32 & & --0.29$\pm$0.04 & 45000 & $\geq$50  & 9 giant & EW & \citet{bragaglia2008}   \\
         & & --0.30$\pm$0.02 & 28000 & $\sim$100 & 2 giant & EW\footnotemark[1]  & \citet{friel2010} \\
Be~39 & & --0.21$\pm$0.01 & 28000 & 70--115 & 3 giant & EW\footnotemark[1]  & \citet{friel2010} \\
Be~66 & & --0.48$\pm$0.24 & 34000 & 5--15     & 2 K giant & EW & \citet{villanova2005}   \\
Be~73 & & --0.22$\pm$0.10 & 40000 & 25--40 & 2 giant & EW & \citet{carraro2007b}   \\
Be~75 & & --0.22$\pm$0.20 & 40000 & 25--40    & 1 giant & EW & \citet{carraro2007b}   \\
Blanco~1 & &  +0.04$\pm$0.02 & 50000 & $\geq$70  & 8 F-G dwarf & Syn & \citet{ford2005}   \\
         & &  +0.20$\pm$0.03 & 28000 & 100--400 & 4 dwarf & EW & \citet{edvardsson1995} \\
Cr~121 & &  +0.25 & 20000 &  & 1 supergiant & EW  & \citet{malllik1998} \\
Cr~261 & & --0.22$\pm$0.11 & 25000 & $\geq$75 & 4 giant & EW & \citet{friel2003}   \\
         & & --0.03$\pm$0.04 & 40000 & 70--130 & 6 giant & EW & \citet{carretta2005}   \\
         & & --0.01$\pm$0.02 & 47000 & 80--100 & 12 giant & EW & \citet{desilva2007}   \\
         & &  +0.13$\pm$0.05 & 45000 & $\geq$60  & 7 giant & EW & \citet{sestito2008} \\ 
Hyades & Mel~25 &  +0.13$\pm$0.02 & 45000 & 200 & 14 F dwarf & EW & \citet{boesgaard1990} \\
         & & --0.05$\pm$0l03 & 60000 & $\sim$200 & 29 F 19 A dwarf & Syn  & \citet{varenne1999} \\
         & &  +0.12$\pm$0.06 & 40000 & 100--300 & 3 giant & EW & \citet{boyarchuk2000} \\ 
         & &  +0.13$\pm$0.08 & 40000 & $\sim$100 & 2 F-K dwarf & EW & \citet{sestito2003}   \\ 
         & &  +0.13$\pm$0.06 & 60000 & 100--200 & 55 F-M dwarf & EW & \citet{paulson2003}   \\
         & &  +0.13$\pm$0.05 & 60000 & 100--200 & 46 F-K dwarf & EW/Syn & \citet{desilva2006}\\
         & &  +0.14$\pm$0.04 & 40000 & 100 & 1 dwarf & EW & \citet{dorazirandich2009} \\ 
         & &  +0.21$\pm$0.04 & 60000 & 175--225 & 3 G dwarf/4 giant & Syn  & \citet{schuler2009,schuler2006} \\
IC~2391 & & --0.03$\pm$0.07 & 18000--44000 & 30--100 & 4 dwarf & EW & \citet{randich2001}   \\
         & & --0.01$\pm$0.02 & 40000 & 70--280 & 7 G--K dwarf & EW  & \citet{dorazirandich2009}\\
IC~2581 & & --0.34$\pm$0.02 & 18000 & $\geq$100 & 1 F supergiant & EW  & \citet{luck1994} \\
IC~2602 & & --0.05$\pm$0.05 & 18000--44000 & 30--100 & 9 dwarf & EW & \citet{randich2001}   \\
         & &  +0.00$\pm$0.01 & 40000 & 100--250 & 8 G--K dwarf & EW & \citet{dorazirandich2009} \\
IC~2714 & &  +0.12$\pm$0.09 & 48000 & 180 & 1 giant & EW\footnote[2]{Spectral synthesis for C, N, O.} & \citet{smiljanic2009}   \\
         & & --0.01$\pm$0.01 & 50000 & 200--300 & 3 giant & EW & \citet{santos2009} \\
IC~4651 & &  +0.11$\pm$0.01 & 40000 & $\geq$100 & 5 giant & EW & \citet{carretta2004}   \\
         & &  +0.10$\pm$0.03 & 45000 & 70--120 & 5 giant/18 dwarf & EW & \citet{pasquini2004}   \\
         & &  +0.12$\pm$0.05 & 100000 & $\sim$80 & 5 G dwarf & EW & \citet{pace2008}  \\
         & &  +0.15$\pm$0.01 & 50000 & 200--300 & 3 giant/3 dwarf & EW  & \citet{santos2009} \\
         & &  +0.11$\pm$0.01 & 48000 & $\geq$100 & 5 giant & EW/Syn  & \citet{mikolaitis2011} \\ 
IC~4665 & & --0.03$\pm$0.04 & 60000 & 30--150 & 18 F--K dwarf & Syn & \citet{shen2005} \\
IC~4725 & M~25 &  +0.18$\pm$0.08 & 18000 & $\geq$100 & 2 supergiant/1 giant & EW  & \citet{luck1994} \\
IC~4756 & & --0.02$\pm$0.05 & 18000 & $\geq$100 & 4 supergiant & EW  & \citet{luck1994} \\
         &  & --0.15$\pm$0.04 & 15000 & 70--150 & 7 giant & EW\footnote[3]{Spectral synthesis for Al, Na, O.} & \citet{jacobson2007}   \\
         & &  +0.04$\pm$0.04 & 48000 & $\geq$170 & 5 giant & EW\footnotemark[2] & \citet{smiljanic2009}   \\
         & &  +0.02$\pm$0.03 & 50000 & 200--300 & 3 giant/3 dwarf & EW  & \citet{santos2009} \\
         & &  +0.01$\pm$0.09 & 100000 & 50--100 & 3 dwarf & EW & \citet{pace2010} \\
         & &  +0.08$\pm$0.11 & 100000 & 50-100 & 3 giant & EW & \citet{pace2010} \\
M~11 & NGC~6705 &  +0.10$\pm$0.07 & 38000 & 85--130 & 10 K giant & EW/Syn & \citet{gonzalez2000}\\
M~34 & NGC~1039 &  +0.07$\pm$0.04 & 45000 & $\sim$70 & 9 G dwarf & EW  & \citet{schuler2003}   \\
M~67 & NGC~2682 &  +0.02$\pm$0.05 & 28000 & 20--50 & 4 F dwarf & EW & \citet{friel1992}\\
       & & --0.03$\pm$0.03 & 30000 & $\geq$100 & 9 giant & EW\footnote[4]{Spectral synthesis for C, N, Eu} & \citet{tautvaisiene2000}  \\
         & & --0.01$\pm$0.04 & 28000 & $\geq$200 & 3 giant & EW & \citet{yong2005}  \\
         & &  +0.03$\pm$0.01 & 45000 & 80--180 & 8 dwarf/2subgiant & EW  & \citet{randich2006}   \\
         & &  +0.03$\pm$0.04 & 100000 & $\sim$80 & 6 G dwarf & EW & \citet{pace2008}  \\
         & &  +0.00$\pm$0.02 & 50000 & 200-300 & 3 giant/3 dwarf & EW & \citet{santos2009}\\
         & &  +0.03$\pm$0.07 & 28000 & 150--180 & 3 giant & EW\footnotemark[1] & \citet{friel2010} \\
         & & --0.01$\pm$0.05 & 21000 & $>$70 & 19 giants & EW\footnotemark[1] & \citet{jacobson2011b} \\
Mel~20 & Alpha &  +0.23$\pm$0.08 & 45000 & 300--450 & 1 supergiant/1 dwarf & EW  & \citet{gonzalez1996}\\
         & persei & --0.05$\pm$0.05 & 45000 & 100 & 7 F dwarf & EW  & \citet{boesgaard1990}\\	
Mel~66 & Cr~147 & --0.38$\pm$0.01 & 30000 & $\sim$100 & 2 giant & EW  & \citet{gratton1994}   \\
         &           & --0.33$\pm$0.03 & 45000 & 80--115  & 5 giant & EW & \citet{sestito2008} \\
Mel~71 &  Cr~155 & --0.30$\pm$0.01 & 34000 & $\sim$100 & 2 giant & EW  & \citet{brown1996}   \\
Mel~111 & Coma & --0.05$\pm$0.05 & 28000 & $\geq$150 & 14 F dwarf & EW  & \citet{friel1992}\\
         & Berenice &  +0.06$\pm$0.10 & 42000 & 150--400 & 11 A 11 F dwarf & Syn  & \citet{gebran2008} \\
NGC~188  &           &  +0.01$\pm$0.08 & 35000 & 20--35 & 11 G dwarf & EW & \citet{randich2003}   \\
         &          &  +0.12$\pm$0.02 & 28000 & 120--140 & 4 giant & EW\footnotemark[1] & \citet{friel2010} \\
         & & --0.03$\pm$0.04 & 21000 & $>$70 & 27 giants & EW\footnotemark[1] & \citet{jacobson2011b} \\
NGC~752  &          &  --0.09$\pm$0.05 & 40000 & 80--150 & 8 dwarf & EW & \citet{hobbs1992} \\
         &        & +0.01$\pm$0.04 & 57000 & 30--80 & 18 G dwarf & EW & \citet{sestito2004} \\
NGC~1193 &       & --0.30$\pm$0.06 & 28000 & 100 & 1 giant & EW\footnotemark[1] & \citet{friel2010} \\
NGC~1245 &  & --0.04$\pm$0.05 & 21000 & $>$70 & 13 giants & EW\footnotemark[1] & \citet{jacobson2011b} \\
NGC~1817 &       & --0.07$\pm$0.04 & 28000 & 120 & 2 giant & EW\footnotemark[1]  & \citet{jacobson2009}   \\
         & & --0.16$\pm$0.03 & 21000 & $>$70 & 28 giants & EW\footnotemark[1] & \citet{jacobson2011b} \\
NGC~1883 &       & --0.20$\pm$0.22 & 20000 & $\sim$20 & 5 giant & EW  & \citet{villanova2007}   \\
         &      & --0.01$\pm$0.01 & 28000 & $\sim$100 & 3 giant & EW\footnotemark[1]  & \citet{jacobson2009}   \\
NGC~1901 &        & --0.08$\pm$0.01 & 33000-64000 & 50--80 & 1 subgiant & EW\footnotemark[1] & \citet{carraro2007} \\
NGC~2112 &        & --0.09$\pm$0.10 & 16000--34000 & 80 & 2 giant & EW & \citet{brown1996}   \\
         &         &  +0.16$\pm$0.03 & 33000 & 80--100 & 3 giant & EW & \citet{carraro2008} \\
NGC~2141 &         & --0.26 & 28000 & $\geq$130 & 1 giant & EW & \citet{yong2005} \\
         &         &  +0.00$\pm$0.16 & 28000 & 75 & 1 giant & EW\footnotemark[1] & \citet{jacobson2009}  \\
NGC~2158 &         & --0.03$\pm$0.14 & 28000 & 75 & 1 giant & EW\footnotemark[1] & \citet{jacobson2009}  \\
         & & --0.28$\pm$0.05 & 21000 & $>$70 & 15 giants & EW\footnotemark[1] & \citet{jacobson2011b} \\
NGC~2194 &   & --0.08$\pm$0.08 & 21000 & $>$70 & 6 giants & EW\footnotemark[1] & \citet{jacobson2011b} \\
NGC~2204 &         & --0.23$\pm$004 & 20000 & $\geq$60 & 13 giant & EW  & \citet{jacobson2011} \\
NGC~2232 &         &  +0.22$\pm$0.09\footnote[5]{Depending on the adopted temperature scale.} & 16000 & 70--300 & 5 dwarf & EW & \citet{monroe2010} \\
        &   & +0.32$\pm$0.08\footnotemark[5] & 16000 & 70--300 & 5 dwarf & EW & \citet{monroe2010} \\
NGC~2243 &         & --0.48$\pm$0.01 & 30000 & $\sim$100 & 2 giant & EW & \citet{gratton1994}  \\
         &         & --0.42$\pm$0.05 & 20000 & $\geq$60 & 10 giant & EW  & \citet{jacobson2011}\\
NGC~2264 &         & --0.18$\pm$0.08 & 45000 & 45--75 & 4 dwarf & EW & \citet{king2000} \\
NGC~2324 &         & --0.17$\pm$0.05 & 45000 & $\geq$80 & 7 giant & EW & \citet{bragaglia2008}  \\
NGC~2355 & & --0.08$\pm$0.08 & 21000 & $>$70 & 5 giants & EW\footnotemark[1] & \citet{jacobson2011b} \\
NGC~2360 &         &  +0.07$\pm$0.06 & 28000 & 50--200 & 4 giant & EW & \citet{hamdani2000}  \\
         &         &  +0.04$\pm$0.09 & 28000 & $\geq$70 & 4 giant & EW\footnotemark[2] & \citet{smiljanic2009}  \\
         &         & --0.03$\pm$0.01 & 50000 & 200--300 & 3 giant & EW & \citet{santos2009}\\ 
NGC~2420 &         & --0.57$\pm$0.08 & 20000 &  & 4 giant & Syn & \citet{smith1987}  \\
         & & --0.20$\pm$0.06 & 21000 & $>$70 & 9 giants & EW\footnotemark[1] & \citet{jacobson2011b} \\
NGC~2423 &         &  +0.14$\pm$0.06 & 50000 & 200--300 & 3 giant & EW  & \citet{santos2009}\\
NGC~2425 &         & --0.15$\pm$0.09 & 21000 & $>$70 & 4 giants & EW\footnotemark[1] & \citet{jacobson2011b} \\
NGC~2447 & M~93 &  +0.03$\pm$0.03 & 28000 & 50--2000 & 3 giant & EW & \citet{hamdani2000}  \\
         &         & --0.01$\pm$0.01 & 28000 & $\geq$70 & 3 giant & EW\footnotemark[2] & \citet{smiljanic2009}\\
         &         & --0.10$\pm$0.03 & 50000 & 100--300 & 3 giant/3 dwarf & EW & \citet{santos2009}\\
NGC~2477 &         &  +0.07$\pm$0.03 & 45000 & $\geq$80 & 6 giant & EW & \citet{bragaglia2008}  \\
NGC~2506 &          & --0.20$\pm$0.02 & 40000 & $\geq$35 & 4 giant & EW & \citet{carretta2004}  \\
         &          & --0.24$\pm$0.05 & 48000 & $\geq$35 & 4 giant & EW & \citet{mikolatis2011b}  \\
NGC~2516 &          &  +0.01$\pm$0.07 & 47000 & 70 & 2 F dwarf & EW & \citet{terndrup2002}\\
NGC~2539 &          &  +0.13$\pm$0.09 & 50000 & 200--300 & 3 giant & EW & \citet{santos2009}\\
NGC~2660 &         &  +0.04$\pm$0.04 & 45000 & $\geq$45 & 5 giant & EW & \citet{bragaglia2008}  \\
NGC~3114 &         &  +0.05$\pm$0.13 & 48000 &  & 7 giant & EW & \citet{pereira2010} \\
         &         &  +0.02$\pm$0.09 & 50000 & 200--300 & 3 giant & EW & \citet{santos2009} \\
NGC~3532 &         &  +0.10$\pm$0.17 & 18000 & $\geq$100 & 5 giant/1 supergiant & EW  & \citet{luck1994} \\
         &         &  +0.04$\pm$0.05 & 48000 & $\geq$170 & 6 giant & EW\footnotemark[2] & \citet{smiljanic2009}  \\
NGC~3680 &         & --0.04$\pm$0.03 &100000 & $\sim$80 & 2 G dwarf & EW & \citet{pace2008} \\
         &         &  +0.04$\pm$0.10 & 48000 & 200 & 1 giant & EW\footnotemark[2] & \citet{smiljanic2009}  \\
         &         & --0.03$\pm$0.01 & 50000 & 200--300 & 3 giant/3 dwarf & EW & \citet{santos2009}\\
NGC~3960 &         &  +0.02$\pm$0.04 & 45000 & $\geq$95 & 6 giant & EW & \citet{bragaglia2008}  \\
NGC~4349 &         & --0.12$\pm$0.06 & 50000 & 200--300 & 3 giant & EW & \citet{santos2009}\\
NGC~5460 &         &  +0.05$\pm$0.24 & 25000 & $\geq$100 & 21 A B F & Syn & \citet{fossati2010} \\
NGC~5822 &         & +0.07$\pm$0.02 & 18000 & $\geq$100 & 1 giant & EW & \citet{luck1994} \\
         & &  +0.04$\pm$0.08 & 48000 & $\geq$130 & 5 giant & EW\footnotemark[2] & \citet{smiljanic2009}  \\
         & &  +0.05$\pm$0.04 & 50000 & 200--300 & 3 giant/3 dwarf & EW & \citet{santos2009}\\
         & &  +0.05$\pm$0.09 & 100000 & 50--100 & 2 dwarf & EW & \citet{pace2010} \\
         & &  +0.15$\pm$0.11 & 100000 & 50--100 & 3 giant & EW & \citet{pace2010} \\
NGC~6067 & &  +0.02$\pm$0.12 & 18000 & $\geq$100 & 1 supergiant & EW  & \citet{luck1994} \\
NGC~6087 & &  +0.06$\pm$0.20 & 18000 & $\geq$100 & 2 supergiant/1 giant & EW  & \citet{luck1994} \\
NGC~6134 & &  +0.15$\pm$0.07 & 40000 & $\geq$60  & 6 gint & EW & \citet{carretta2004}  \\
         & &  +0.12$\pm$0.09 & 48000 & $\geq$150 & 3 giant & EW\footnotemark[2] & \citet{smiljanic2009}  \\
         & &  +0.15$\pm$0.05 & 45000 & $\geq$60 & 6 giant & EW\footnotemark[1]  & \citet{mikilaitis2010} \\
NGC~6192 & &  +0.12$\pm$0.04 & 47000 & $\geq$140 & 4 giant & EW  & \citet{magrini2010} \\
NGC~6253 & &  +0.46$\pm$0.03 & 48000 & 85--180 & 5 giant & Syn & \citet{carretta2007b} \\
         & &  +0.36$\pm$0.07 & 47000 & 65--140 & 2 giant/1 subgiant/1 dwarf & EW & \citet{sestito2007}  \\
NGC~6281 & &  +0.05$\pm$0.06 & 48000 & $\geq$230 & 2 giant & EW\footnotemark[2] & \citet{smiljanic2009}  \\
NGC~6404 & &  +0.11$\pm$0.04 & 47000 & $\geq$110 & 4 giant & EW  & \citet{magrini2010}  \\
NGC~6475 & M~7 &  +0.14$\pm$0.06 & 48000 & 50--150 & 13 F-K dwarf & EW & \citet{sestito2003}  \\
         & &  +0.03$\pm$0.02 & 18000 & $\geq$200 & 2 B 3 F--K dwarf/2 G--K giant & EW & \citet{villanova2009} \\
NGC~6583 & &  +0.37$\pm$0.03 & 47000 & $\geq$80 & 2 giant & EW  & \citet{magrini2010}\\ 
NGC~6633 & &  +0.07$\pm$0.05 & 48000 & $\geq$160 & 2 giant & EW\footnotemark[2] & \citet{smiljanic2009}  \\
         & &  +0.06$\pm$0.01 & 50000 & 200--300 & 3 giant & EW & \citet{santos2009} \\
NGC~6791 & &  +0.37$\pm$0.03 & 20000 & 30 & 1 hot HB & EW/Syn  & \citet{peterson1998}  \\
         & &  +0.35$\pm$0.02 & 25000 & $\geq$40 & 6 giant & Syn  & \citet{origlia2006}  \\
         & &  +0.38$\pm$0.02 & 20000 & 30--60 & 10 giant & Syn  & \citet{carraro2006} \\
         & &  +0.47$\pm$0.07 & 29000 & 40--85 & 4 giant & Syn & \citet{carretta2007b} \\
         & &  +0.30$\pm$0.08 & 45000 & $\sim$40 & 2 dwarf & EW  & \citet{boesgaard2009}  \\
NGC~6819 & &  +0.09$\pm$0.03 & 40000 & 130 & 3 giant & EW & \citet{bragaglia2001}  \\
NGC~6882/5 & & --0.03$\pm$0.01 & 18000 & $\geq$100 & 1 supergiant/1 dwarf & EW & \citet{luck1994} \\
NGC~6939 & &  +0.00$\pm$0.10 & 15000 & 70--150 & 9 giant & EW\footnotemark[3] & \citet{jacobson2007}  \\
NGC~7142 & & +0.08$\pm$0.06 & 15000 & 70--150 & 6 giant & EW\footnotemark[3] & \citet{jacobson2007}\\
         & &  +0.16$\pm$0.01 & 30000 & 100--130 & 4 giant & EW\footnotemark[1] & \citet{jacobson2008}  \\
NGC~7160 & & +0.16$\pm$0.03 & 16000 & 70--300 & 16 F--G dwarf & EW & \citet{monroe2010}  \\
NGC~7789 & & --0.04$\pm$0.05 & 30000 & $\geq$50 & 9 giant & EW/Syn  & \citet{tautvaisiene2005} \\
         & & +0.02$\pm$0.04 & 21000 & $>$70 & 28 giants & EW\footnotemark[1] & \citet{jacobson2011b} \\
Pleiades & M~45 & --0.03$\pm$0.02 & 45000 & 150 & 12 F dwarf & EW  & \citet{boesgaard1990}\\
         & & --0.03$\pm$0.10 & 18000--44000 & 30--100 & 2 dwarf & EW & \citet{randich2001}  \\
         & &  +0.06$\pm$0.02 & 42000--75000 & 100--300 & 16 A giant/5 F dwarf & Syn&\citet{grebranmonier2008}\\
         & &  +0.07$\pm$0.03 & 45000 & 70 & 2 dwarf & EW  & \citet{king2000} \\
         & &  +0.07$\pm$0.05 & 40000 &  & 20 F--G--K dwarf & Syn & \citet{soderblom2009}\\
Praesepe & M~44 &  +0.04$\pm$0.04 & 28000 & 60--190 & 6 F dwarf & EW  & \citet{friel1992}  \\
         & Mel~88 &  +0.27$\pm$0.04 & 100000 & $\sim$130 & 7 G dwarf & EW & \citet{pace2008} \\
Rup~4 &         & --0.09$\pm$0.04 & 40000 & 25--40   & 3 giant & EW & \citet{carraro2007b}  \\
Rup~7 &  & --0.26$\pm$0.05 & 40000 & 25--40   & 5 giant & EW & \citet{carraro2007b}  \\
Saurer~1 & & --0.38$\pm$0.01 & 34000 & $\sim$80 & 2 G giant & EW & \citet{carraro2004} \\
To~2 & & --0.50$\pm$0.10 & 34000 & 50--60 & 3 giant & EW  & \citet{brown1996}  \\
         & & --0.28$\pm$0.01 & 21000--40000 & 15--70 & 18 giant & EW & \citet{frinchaboy2008} \\
         & & --0.31$\pm$0.02 & 17000 & 60--80 & 13 giant & EW & \citet{villanova2010} \\	 
Tr~20 & & --0.11$\pm$0.13 & 50000 & 65 & 1 giant & EW & \citet{platais2008}\\
\end{longtable}
\end{landscape}
}

\begin{thebibliography}{}

\bibitem[Alonso et al.(1999)]{alonso99} Alonso, A., Arribas, S., \& Mart{\'{\i}}nez-Roger, C.\ 1999, A\&AS, 140, 261 

\bibitem[An et al.(2007)]{an2007} An, D., Terndrup, D.~M., 
Pinsonneault, M.~H., Paulson, D.~B., Hanson, R.~B., 
\& Stauffer, J.~R.\ 2007, \apj, 655, 233 

\bibitem[Andersen(1999)]{andersen1999} Andersen, J.\ 1999,  Transactions of the IAU, Vol.~XXIVA, p.36 (1999) , 24, A36 

\bibitem[Andreuzzi et al.(2011)]{andreuzzi2011} Andreuzzi, G., 
Bragaglia, A., Tosi, M., \& Marconi, G.\ 2011, \mnras, 412, 1265 

\bibitem[Andrievsky et al.(2004)]{andrievsky2004} Andrievsky, S.~M., Luck, R.~E., Martin, P., \& L{\'e}pine, J.~R.~D.\ 2004, \aap, 413, 159 


\bibitem[Arp(1962)]{arp1962} Arp, H.\ 1962, \apj, 136, 66 

\bibitem[Barry et al.(1987)]{barry1987} Barry, D.~C., Cromwell, R.~H., \& Hege, E.~K.\ 1987, \apj, 315, 264 

\bibitem[Barta{\v s}i{\= u}t{\.e} et al.(2003)]{bartasiute2003} 
Barta{\v s}i{\= u}t{\.e}, S., Aslan, Z., Boyle, R.~P., Kharchenko, N.~V., 
Ossipkov, L.~P., \& Sperauskas, J.\ 2003, Baltic Astronomy, 12, 539 

\bibitem[Barta{\v s}i{\= u}t{\.e} et al.(2007)]{bartasiute2007} Barta{\v s}i{\= u}t{\.e}, S., Deveikis, V., Strai{\v z}ys, V., 
\& Bogdanovi{\v c}ius, A.\ 2007, Baltic Astronomy, 16, 199 

\bibitem[Bensby et al.(2004)]{bensby2004} Bensby, T., Feltzing, S., \& Lundstr{\"o}m,
I.\ 2004, \aap, 421, 969 

\bibitem[Bessell(1979)]{bessell1979} Bessell, M.~S.\ 1979, PASP,  91, 589 

\bibitem[Blake(2002)]{blake2002} Blake, R.~M.\ 2002, Ph.D.~Thesis,  

\bibitem[Blake \& Rucinski(2004)]{blake2004} Blake, R.~M., \& Rucinski, S.~M.\ 2004, Bulletin of the American Astronomical Society, 36, 1483 

\bibitem[Boesgaard(1989)]{boesgaard1989} Boesgaard, A.~M.\ 1989, 
\apj, 336, 798 

\bibitem[Boesgaard \& Friel(1990)]{boesgaard1990} Boesgaard, A.~M., \& Friel, E.~D.\ 1990, \apj, 35

\bibitem[Boesgaard(1991)]{boesgaard1991} Boesgaard, A.~M.\ 1991, \apjl, 370, L95 

\bibitem[Boesgaard et al.(2009)]{boesgaard2009} Boesgaard, A.~M.,  Jensen, E.~E.~C., \&
Deliyannis, C.~P.\ 2009, \aj, 137, 4949 

\bibitem[Bouvier et al.(2008)]{bouvier2008} Bouvier, J., et al.\ 2008, \aap, 481, 661

\bibitem[Boyarchuk et al.(2000)]{boyarchuk2000} Boyarchuk, A.~A., Antipova, L.~I., Boyarchuk, M.~E., \& Savanov, I.~S.\ 2000, Astronomy Reports, 44, 76

\bibitem[Bragaglia et al.(2001)]{bragaglia2001} Bragaglia, A., et  al.\ 2001, \aj, 121,
327 

\bibitem[Bragaglia et al.(2008)]{bragaglia2008} Bragaglia, A., Sestito, P., Villanova, S., Carretta, E., Randich, S., \& Tosi, M.\ 2008, \aap, 480, 79 

\bibitem[Brown et al.(1996)]{brown1996} Brown, J.~A.,  Wallerstein, G., Geisler, D.,
\& Oke, J.~B.\ 1996, \aj, 112, 1551 

\bibitem[Burkhart \& Coupry(1998)]{burkhart1998} Burkhart, C., \& Coupry, M.~F.\ 1998, \aap, 338, 1073 

\bibitem[Burkhart \& Coupry(2000)]{burkhart2000} Burkhart, C., \& Coupry, M.~F.\ 2000, \aap, 354, 216 

\bibitem[Cardelli et al.(1989)]{cardelli89} Cardelli, J.~A., Clayton, G.~C., \& Mathis, J.~S.\ 1989, ApJ, 345, 245 

\bibitem[Carraro et al.(1993)]{carraro1993} Carraro, G., Bertelli, G., Bressan, A., \& Chiosi, C.\ 1993, \aaps, 101, 381 

\bibitem[Carraro \& Chiosi(1994)]{carraro1994} Carraro, G., \& Chiosi, C.\ 1994, \aap, 287, 761 

\bibitem[Carraro et al.(1998)]{carraro1998} Carraro, G., Ng, Y.~K., 
\& Portinari, L.\ 1998, \mnras, 296, 1045 

\bibitem[Carraro et al.(2004)]{carraro2004} Carraro, G., Bresolin,  F., Villanova, S.,
Matteucci, F., Patat, F.,  \& Romaniello, M.\ 2004, \aj, 128, 1676 

\bibitem[Carraro et al.(2006)]{carraro2006} Carraro, G., Villanova,  S., Demarque, P.,
McSwain, M.~V., Piotto, G.,  \& Bedin, L.~R.\ 2006, \apj, 643, 1151 

\bibitem[Carraro et al.(2007a)]{carraro2007} Carraro, G., de La Fuente Marcos, R., Villanova, S., Moni Bidin, C., de La Fuente Marcos, C., Baumgardt, H., \& Solivella, G.\ 2007a, \aap, 466, 931 

\bibitem[Carraro et  al.(2007b)]{carraro2007b} Carraro, G., Geisler, D., Villanova, S.,
Frinchaboy, P.~M., \& Majewski, S.~R.\ 2007b, \aap, 476, 217 

\bibitem[Carraro et al.(2008)]{carraro2008} Carraro, G., Villanova, 
S., Demarque, P., Moni Bidin, C., \& McSwain, M.~V.\ 2008, \mnras, 386, 1625 

\bibitem[Carraro \& Bensby(2009)]{carraro2009} Carraro, G., \& Bensby, T.\ 2009, \mnras, 397, L106 


\bibitem[Carretta et al.(2004)]{carretta2004} Carretta, E., Bragaglia, A., Gratton, R.~G., \& Tosi, M.\ 2004, \aap, 422, 951 

\bibitem[Carretta et  al.(2005)]{carretta2005} Carretta, E., Bragaglia, A., Gratton,
R.~G., \& Tosi, M.\ 2005, \aap, 441, 131 



\bibitem[Carretta et  al.(2007)]{carretta2007b} Carretta, E., Bragaglia, A., \& Gratton,
R.~G.\ 2007, \aap, 473, 129 

\bibitem[Carretta et al.(2010)]{carretta2010} Carretta, E., Bragaglia, A., Gratton, R.~G., Recio-Blanco, A., Lucatello, S., D'Orazi, V., \& Cassisi, S.\ 2010, \aap, 516, A55 

\bibitem[Cayrel et al.(2004)]{cayrel2004} Cayrel, R., et al.\ 2004, \aap, 416, 1117

\bibitem[Chen et al.(2003)]{chen2003}Chen , L., Hou, J. L. \& Wang, J. J.\ 2003, AJ, 125, 1397

\bibitem[Chiappini et al.(2001)]{chiappini2001} Chiappini, C., 
Matteucci, F., \& Romano, D.\ 2001, \apj, 554, 1044 

\bibitem[Claria et al.(1996)]{claria1996} Claria, J.~J., Piatti, A.~E., \& Osborn, W.\ 1996, \pasp, 108, 672 


\bibitem[Coleman(1982)]{coleman82} Coleman, L.~A.\ 1982, \aj, 87, 369 

\bibitem[Costa et al.(2004)]{costa2004} Costa, R.~D.~D., Uchida, M.~M.~M., \& Maciel, W.~J.\ 2004, \aap, 423, 199 

\bibitem[Crawford \& Barnes(1970)]{crawford1970} Crawford, D.~L., \& Barnes, J.~V.\ 1970, \aj, 75, 946 

\bibitem[Cutri et al.(2003)]{2mass} Cutri, R.M., Skrutskie, M.F., Van Dyk, et al.\ 2003, Expla\-na\-tory Supplement to the 2MASS All
Sky Data Release
\bibitem[Daflon \& Cunha(2004)]{daflon2004} Daflon, S., \& Cunha, K.\ 2004, \apj, 617, 1115 

\bibitem[Daniel et al.(1994)]{daniel1994} Daniel, S.~A., Latham, D.~W., Mathieu, R.~D., \& Twarog, B.~A.\ 1994, \pasp, 106, 281 

\bibitem[Dean et al.(1978)]{dean1978} Dean, J.~F., Warren,  P.~R., \& Cousins, A.~W.~J.\ 1978, MNRAS, 183, 569 

\bibitem[Deharveng et al.(2000)]{deharveng2000} Deharveng, L., Pe{\~n}a, M., Caplan, J., \& Costero, R.\ 2000, \mnras, 311, 329 

\bibitem[Den Hartog et al.(2003)]{denhartog2003} Den Hartog, E. A., Lawler, J. E.,Sneden, C., Cowan, J. J.\ 2003, ApJS, 148, 543

\bibitem[De Silva et al.(2006)]{desilva2006} De Silva, G.~M., Sneden, C., Paulson, D.~B., Asplund, M., Bland-Hawthorn, J., Bessell, 
M.~S., \& Freeman, K.~C.\ 2006, \aj, 131, 455 

\bibitem[De Silva et al.(2007)]{desilva2007} De Silva, G.~M.,  Freeman, K.~C.,
Asplund, M., Bland-Hawthorn, J., Bessell, M.~S.,  \& Collet, R.\ 2007, \aj, 133,
1161 

\bibitem[de Silva et al.(2009)]{desilva2009} de Silva, G.~M., Gibson, B.~K.,
Lattanzio, J., \& Asplund, M.\ 2009, \aap, 500, L25 

\bibitem[Dias et al.(2002)]{dias2002}Dias, W.~S., Alessi,  B.~S., Moitinho, A., \& L{\'e}pine, J.~R.~D.\ 2002, \aap, 389, 871 

\bibitem[Dinescu et al.(1995)]{dinescu1995} Dinescu, D.~I., Demarque, P., Guenther, D.~B., \& Pinsonneault, M.~H.\ 1995, \aj, 109, 2090 

\bibitem[D'Orazi et al.(2006)]{dorazi2006} D'Orazi, V., Bragaglia, A., Tosi, M., Di Fabrizio, L., \& Held, E.~V.\ 2006, \mnras, 368, 471 

\bibitem[D'Orazi et al.(2009)]{dorazi2009} D'Orazi, V., Magrini,  L., Randich, S.,
Galli, D., Busso, M.,  \& Sestito, P.\ 2009, \apjl, 693, L31 

\bibitem[D'Orazi \& Randich(2009)]{dorazirandich2009} D'Orazi, V., \& Randich, S.\ 2009, \aap, 501, 553 

\bibitem[Dutra \& Bica(2000)]{dutra2000} Dutra, C.~M., \& Bica, E.\ 2000, \aap, 359, 347 

\bibitem[Dzervitis \& Paupers(1993)]{dzervitis1993} Dzervitis, U., \& Paupers, O.\ 1993, \apss, 199, 77 



\bibitem[Edvardsson et al.(1993)]{edvardson1993} Edvardsson, B., Andersen, J., Gustafsson, B., Lambert, D. L., Nissen, P. E., Tomkin, J., \ 1993,
A\&A, 275, 101

\bibitem[Edvardsson et al.(1995)]{edvardsson1995} Edvardsson, B., Pettersson, B., Kharrazi, M., \& Westerlund, B.\ 1995, \aap, 293, 75
 
\bibitem[Eggen(1963)]{eggen1963} Eggen, O.~J.\ 1963, \apj, 138, 
356 

\bibitem[Eggen(1989)]{eggen1989} Eggen, O.~J.\ 1989, \pasp, 101, 
54 
\bibitem[Eggen(1998)]{eggen1998} Eggen, O.~J.\ 1998, \aj, 116, 284 

\bibitem[Famaey et al.(2005)]{Famaey2005} Famaey, B., Jorissen, A., Luri, X., Mayor, M., Udry, S., Dejonghe, H., \& Turon, C.\ 2005, \aap, 430, 165 

\bibitem[Famaey et al.(2007)]{famaey2007} Famaey, B., Pont, F., Luri, X., Udry, S., Mayor, M., \& Jorissen, A.\ 2007, \aap, 461, 957 
\bibitem[Feltzing et al.(2001)]{feltzing2001} Feltzing, S., Holmberg, J., \& Hurley, J.~R.\ 2001, \aap, 377, 911 

\bibitem[Ford et al.(2005)]{ford2005} Ford, A., Jeffries, R.~D.,  \& Smalley, B.\
2005, \mnras, 364, 272 

\bibitem[Fossati et al.(2011)]{fossati2010} Fossati, L., Folsom, 
C.~P., Bagnulo, S., Grunhut, J.~H., Kochukhov, O., Landstreet, J.~D., 
Paladini, C., \& Wade, G.~A.\ 2011, \mnras, 413, 1132 

\bibitem[Francic(1989)]{francic1989} Francic, S.~P.\ 1989, \aj, 98, 888 

\bibitem[Friel \& Boesgaard(1992)]{friel1992} Friel, E.~D., \& Boesgaard, A.~M.\ 1992, \apj, 387, 170 

\bibitem[Friel \& Janes(1993)]{friel1993}Friel E.D., Janes K.A. \ 1993 A\&A, 267, 75 

\bibitem[Friel(1995)]{friel1995} Friel, E.~D.\ 1995, \araa, 33, 381 

\bibitem[Friel et al.(2002)]{friel02} Friel, E.~D., Janes, K.~A., Tavarez, M., Scott, J., Katsanis, R., Lotz, J., Hong, L., \& Miller,  N.\ 2002, \aj, 124, 2693 

\bibitem[Friel et al.(2003)]{friel2003} Friel, E.~D., Jacobson,  H.~R., Barrett, E., Fullton, L., Balachandran, S.~C.,  \& Pilachowski, C.~A.\ 2003, \aj, 126, 2372 

\bibitem[Friel et al.(2005)]{friel2005} Friel, E.~D., Jacobson,  H.~R., \& Pilachowski, C.~A.\ 2005, \aj, 129, 2725 

\bibitem[Friel et al.(2010)]{friel2010} Friel, E.~D., Jacobson, 
H.~R., \& Pilachowski, C.~A.\ 2010, \aj, 139, 1942 

\bibitem[Frinchaboy et al.(2008)]{frinchaboy2008} Frinchaboy, P.~M.,  Marino, A.~F.,
Villanova, S., Carraro, G., Majewski, S.~R.,  \& Geisler, D.\ 2008, \mnras, 391,
39 

\bibitem[Fulbright et al.(2007)]{fulbright2007} Fulbright, J.~P., 
McWilliam, A., \& Rich, R.~M.\ 2007, \apj, 661, 1152 
\bibitem[G{\'a}sp{\'a}r et al.(2009)]{gaspar2009} G{\'a}sp{\'a}r, 
A., Rieke, G.~H., Su, K.~Y.~L., Balog, Z., Trilling, D., Muzzerole, J., 
Apai, D., \& Kelly, B.~C.\ 2009, \apj, 697, 1578

\bibitem[Gebran et al.(2008)]{gebran2008} Gebran, M., Monier, R., \& Richard, O.\ 2008, \aap, 479, 189 

\bibitem[Gebran \& Monier(2008)]{grebranmonier2008} Gebran, M., \& Monier, R.\ 2008, \aap,
483, 567 

\bibitem[Giardino et al.(2008)]{giardino2008} Giardino, G., Pillitteri, I., Favata, F., \& Micela, G.\ 2008, \aap, 490, 113

\bibitem[Gilmore et al.(1995)]{gilmore1995} Gilmore, G., Wyse, R.~F.~G., \& Jones, J.~B.\ 1995, \aj, 109, 1095 

\bibitem[Girardi et al.(2000)]{girardi2000} Girardi, L., Mermilliod, J.-C., \& Carraro, G.\ 2000, \aap, 354, 892 

\bibitem[Girardi \& Salaris(2001)]{girardi2001}Girardi, L., \& Salaris,M.\ 2001, \mnras, 323, 109 

\bibitem[Gonzalez \& Lambert(1996)]{gonzalez1996} Gonzalez, G., \& Lambert, D.~L.\ 1996, \aj, 111, 424 

\bibitem[Gonzalez  \& Wallerstein(2000)]{gonzalez2000} Gonzalez, G., \& Wallerstein,
G.\ 2000, \pasp, 112, 1081 

\bibitem[Gratton  \& Contarini(1994)]{gratton1994} Gratton, R.~G., \& Contarini, G.\
1994, \aap, 283, 911 

\bibitem[Gratton et al.(1999)]{gratton1999} Gratton, R. G., Carretta, E., Eriksson, K., Gustafsson, B.\ 1999, A\&A, 350, 955


\bibitem[Grevesse et al.(1996)]{gre96}Grevesse, N., Noels,  A., \&
Sauval, A.~J.\ 1996, ASP Conf.~Ser.~ 99: Cosmic Abundances, 99, 117 

\bibitem[Griffin et al.(1988)]{griffin88} Griffin, R.~F., Griffin, R.~E.~M., Gunn, J.~E., \& Zimmerman, B.~A.\ 1988, \aj, 96, 172 


\bibitem[Hamdani et  al.(2000)]{hamdani2000} Hamdani, S., North, P., Mowlavi, N.,
Raboud, D., \& Mermilliod, J.-C.\ 2000, \aap, 360, 509 

\bibitem[Hardy(1979)]{hardy1979} Hardy, E.\ 1979, \aj, 84, 319

\bibitem[Heinemann(1926)]{Heinemann1926} Heinemann, K.\ 1926,  Astronomische Nachrichten, 227, 193 

\bibitem[Hernandez et al.(1998)]{hernandez1998} Hernandez, M.~M., Perez Hernandez, F., Michel, E., Belmonte, J.~A., Goupil, M.~J., \& Lebreton, Y.\ 1998, \aap, 338, 511 

\bibitem[Hertzsprung(1909)]{herttzsprung1909} Hertzsprung, E.\ 1909, \apj, 30, 135

\bibitem[Hill \& Pasquini(1999)]{hill1999} Hill, V., \& Pasquini, L.\ 1999, \aap, 348, L21 

\bibitem[Hobbs \& Thorburn(1992)]{hobbs1992} Hobbs, L.~M., \& Thorburn, J.~A.\ 1992, \aj, 104, 669 



\bibitem[H{\o}g et al.(2000)]{hog2000} H{\o}g, E., et al.\ 2000, \aap, 355, L27

\bibitem[Hui-Bon-Hoa 
\& Alecian(1998)]{huibonhoa1998} Hui-Bon-Hoa, A., \& Alecian, G.\ 1998, \aap, 332, 224 

\bibitem[Jacobson et al.(2007)]{jacobson2007} Jacobson, H.~R.,  Friel, E.~D., \&
Pilachowski, C.~A.\ 2007, \aj, 134, 1216 

\bibitem[Jacobson et al.(2008)]{jacobson2008} Jacobson, H.~R., 
Friel, E.~D., \& Pilachowski, C.~A.\ 2008, \aj, 135, 2341 

\bibitem[Jacobson et al.(2009)]{jacobson2009} Jacobson, H.~R.,  Friel, E.~D., \&
Pilachowski, C.~A.\ 2009, \aj, 137, 4753 

\bibitem[Jacobson et al.(2011a)]{jacobson2011} Jacobson, H.~R., 
Friel, E.~D., \& Pilachowski, C.~A.\ 2011a, \aj, 141, 58 

\bibitem[Jacobson et al.(2011b)]{jacobson2011b} Jacobson, H.~R., 
Pilachowski, C.~A., \& Friel, E.~D.\ 2011b, \aj, 142, 59 

\bibitem[Jacquinet-Husson et al.(1999)]{geisa1}Jacquinet-Husson, N., Ari, E., Ballard, J., Barbe, A., Bjoraker, G. et al.\ 1999,  JQSRT, 62, 205

\bibitem[Jacquinet-Husson et al.(2005)]{geisa2}Jacquinet-Husson, N., Scott, N.~A., Garceran, K., Armante, R., Ch\'edin, A.\ 2005, JQSRT, 95, 429

\bibitem[Jameson et al.(2008)]{jameson2008} Jameson, R.~F., Lodieu, N., Casewell,
S.~L., Bannister, N.~P., \& Dobbie, P.~D.\ 2008, \mnras, 385, 1771 

\bibitem[Janes(1979)]{janes1979} Janes, K.~A.\ 1979, \apjs, 39, 135 

\bibitem[Jennens \& Helfer(1975)]{jennens75} Jennens, P.~A., \& Helfer, H.~L.\
1975, \mnras, 172, 681

\bibitem[Johansson et al.(2003)]{johansson03} Johansson, S., Litz{\'e}n,  U.,
Lundberg, H., \& Zhang, Z.\ 2003, \apjl, 584, L107 

\bibitem[Johnson(1952)]{johnson1952} Johnson, H.~L.\ 1952, \apj, 
116, 640 

\bibitem[Johnson(1953)]{johnson1953} Johnson, H.~L.\ 1953, \apj, 117, 356 

\bibitem[Johnson \& Knuckles(1955)]{johnson1955} Johnson, H.~L., \& Knuckles, C.~F.\ 1955, \apj, 122, 209 

\bibitem[Johnson(1961)]{johnson1961} Johnson, H.~L.\ 1961, Lowell Observatory Bulletin, 5, 133

\bibitem[Johnson et al.(1966)]{Johnson1966} Johnson, H.~L., Iriarte, B., Mitchell, R.~I., \& Wisniewskj, W.~Z.\ 1966, Communications of the Lunar and Planetary Laboratory, 4, 99 

\bibitem[Kaluzny \& Mazur(1991)]{kaluzny1991} Kaluzny, J., \& Mazur, B.\ 1991, Acta Astronomica, 41, 167 

\bibitem[Karata{\c s} et al.(2005)]{karatas2005} Karata{\c s}, Y., 
Bilir, S., \& Schuster, W.~J.\ 2005, \mnras, 360, 1345 

\bibitem[King et al.(2000)]{king2000} King, J.~R., Soderblom, 
D.~R., Fischer, D., \& Jones, B.~F.\ 2000, \apj, 533, 944 

\bibitem[Klein Wassink(1927)]{kleinWassink1927} Klein Wassink, W.~J.\ 
1927, Publications of the Kapteyn Astronomical Laboratory Groningen, 41, 1 

\bibitem[Komarov \& Basak(1993)]{komarov1993} Komarov, N.~S., \& Basak, N.~Y.\ 1993, \azh, 70, 111 

\bibitem[Kraus \& Hillenbrand(2007)]{kraus2007} Kraus, A.~L., \& Hillenbrand, L.~A.\ 2007, \aj, 134, 2340 

\bibitem[Kupka et al.(1999)]{vald} Kupka, F., Piskunov, N., Ryabchikova, T.~A., Stempels, H.~C., \& Weiss, W.~W.\ 1999, \aaps, 138, 119 


\bibitem[Lata et al.(2002)]{lata2002} Lata, S., Pandey, A.~K., Sagar, R., \& Mohan, V.\ 2002, \aap, 388, 158 

\bibitem[Lemasle et al.(2008)]{lemasle2008} Lemasle, B., Fran{\c c}ois, P., Piersimoni, A., Pedicelli, S., Bono, G., Laney, C.~D., Primas, F., \& Romaniello, M.\ 2008, \aap, 490, 613 

\bibitem[Letarte et al.(2006)]{letarte2006} Letarte, B., Hill, V., Jablonka, P., Tolstoy, E., Fran{\c c}ois, P., \& Meylan, G.\ 2006, \aap, 453, 547 

\bibitem[Loktin(2000)]{loktin2000} Loktin, A.~V.\ 2000, Astronomy Letters, 26, 657 

\bibitem[Loktin \& Beshenov(2001)]{loktin2001} Loktin, A.~V., \& Beshenov, G.~V.\ 2001, Astronomy Letters, 27, 386 

\bibitem[Luck(1994)]{luck1994} Luck, R.~E.\ 1994, \apjs, 91, 309 

\bibitem[Luck \& Challener(1995)]{luck1995} Luck, R.~E., \& Challener, S.~L.\ 1995, \aj, 110, 2968 

\bibitem[Maciel et al.(2003)]{maciel2003} Maciel, W.~J., Costa, R.~D.~D., \& Uchida, M.~M.~M.\ 2003, \aap, 397, 667 

\bibitem[Maciel et al.(2007)]{maciel2007} Maciel, W.~J., Quireza, C., \& Costa, R.~D.~D.\ 2007, \aap, 463, L13 

\bibitem[Maciel \& Costa(2009)]{maciel2009} Maciel, W.~J., \& Costa, R.~D.~D.\ 2009, IAU Symposium, 254, 38P 

\bibitem[Maeder(1971)]{maeder1971} Maeder, A.\ 1971, \aap, 10, 354 

\bibitem[Magain(1984)]{magain1984} Magain, P.\ 1984, \aap, 134,  189 

\bibitem[Magrini et al.(2009)]{magrini2009} Magrini, L., Sestito, P., Randich, S., \& Galli, D.\ 2009, \aap, 494, 95 

\bibitem[Magrini et al.(2010)]{magrini2010} Magrini, L., Randich, 
S., Zoccali, M., Jilkova, L., Carraro, G., Galli, D., Maiorca, E., 
\& Busso, M.\ 2010, \aap, 523, 11

\bibitem[Mallik(1998)]{malllik1998} Mallik, S.~V.\ 1998, \aap, 338, 623 

\bibitem[Marsakov \& Borkova(2005)]{marsakov2005} Marsakov, V.~A., \& Borkova, T.~V.\ 2005, Astronomy Letters, 31, 515 

\bibitem[Marsakov \& Borkova(2006)]{marsakov2006} Marsakov, V.~A., \& Borkova, T.~V.\ 2006, Astronomy Letters, 32, 376 

\bibitem[Martell \& Smith(2009)]{martell2009} Martell, S.~L., \& Smith,
G.~H.\ 2009, \pasp, 121, 577 

\bibitem[Martell \& Grebel(2010)]{martell2010} Martell, S.~L., \&
Grebel, E.~K.\ 2010, \aap, 519, A14 

\bibitem[Mathieu \& Mazeh(1988)]{mathieu1988} Mathieu, R.~D., \& Mazeh, T.\ 1988, \apj, 326, 256

\bibitem[Mazzei \& Pigatto(1988)]{mazzei1988} Mazzei, P., \& Pigatto, L.\ 1988, \aap, 193, 148 

\bibitem[Mendoza(1967)]{mendoza1967} Mendoza, E.~E.\ 1967, Bolet\'{\i}n de los Observatorios Tonantzintla y Tacubaya, 4, 149 

\bibitem[Mermilliod(1995)]{mermilliod1995}Mermilliod, J.-C.\ 1995, ASSL
Vol.~203: Information \& On-Line Data in Astronomy, 127 

\bibitem[Mermilliod et al.(1998)]{mermilliod98} Mermilliod, J.-C., Mathieu, R.~D., Latham, D.~W., \& Mayor, M.\ 1998, \aap, 339, 423 

\bibitem[Mermilliod et al.(2007)]{mermilliod2007} Mermilliod, J.-C., Andersen, J., Latham, D.~W., \& Mayor, M.\ 2007, \aap, 473, 829

\bibitem[Meynet et al.(1993)]{meynet1993} Meynet, G., Mermilliod, J.-C., \& Maeder, A.\ 1993, \aaps, 98, 477 

\bibitem[Mikolaitis et al.(2010)]{mikilaitis2010} Mikolaitis, {\v S}., 
Tautvai{\v s}ien{\.e}, G., Gratton, R., Bragaglia, A., 
\& Carretta, E.\ 2010, \mnras, 407, 1866 

\bibitem[Mikolaitis et al.(2011)]{mikolaitis2011} Mikolaitis, {\v S}., 
Tautvai{\v s}ien{\.e}, G., Gratton, R., Bragaglia, A., 
\& Carretta, E.\ 2011a, \mnras, 413, 2199

\bibitem[Mikolaitis et al.(2011)]{mikolatis2011b} Mikolaitis, {\v S}., 
Tautvai{\v s}ien{\.e}, G., Gratton, R., Bragaglia, A., 
\& Carretta, E.\ 2011b, arXiv:1105.4047 

\bibitem[Milone et al.(1995)]{milone1995} Milone, E.~F., Stagg, C.~R., Sugars, B.~A., McVean, J.~R., Schiller, S.~J., Kallrath, J., 
\& Bradstreet, D.~H.\ 1995, \aj, 109, 359 

\bibitem[Mishenina et al.(2006)]{mishenina2006} Mishenina, T.~V., Bienaym{\'e}, O., Gorbaneva, T.~I., Charbonnel, C., Soubiran, C., Korotin, S.~A., \& Kovtyukh, V.~V.\ 2006, \aap, 456, 1109 

\bibitem[Mishenina et al.(2007)]{mishenina2007} Mishenina, T.~V., 
Gorbaneva, T.~I., Bienaym{\'e}, O., Soubiran, C., Kovtyukh, V.~V., 
\& Orlova, L.~F.\ 2007, Astronomy Reports, 51, 382 


\bibitem[Molla et al.(1996)]{molla1996} Molla, M., Ferrini, F., \& Diaz, A.~I.\ 1996, \apj, 466, 668 

\bibitem[Montegriffo et al.(1998)]{montegriffo98} Montegriffo, P.,  Ferraro, F.~R., Origlia, L., \& Fusi Pecci, F.\ 1998, \mnras, 297, 872

\bibitem[Monroe \& Pilachowski(2010)]{monroe2010} Monroe, T.~R., \& Pilachowski, C.~A.\ 2010, \aj, 140, 2109

\bibitem[Mould(2005)]{mould2005} Mould, J.\ 2005, \aj, 129, 698

\bibitem[Mucciarelli et al.(2009)]{mucciarelli2009} Mucciarelli, A., 
Origlia, L., Ferraro, F.~R., \& Pancino, E.\ 2009, \apjl, 695, L134 

\bibitem[Mucciarelli(2011)]{mucciarelli11} Mucciarelli, A.\ 2011,  \aap, 528, 44

\bibitem[Narayanan \& Gould(1999)]{narayanan1999} Narayanan, V.~K., \& Gould, A.\ 1999, \apj, 515, 256 

\bibitem[Navarro et al.(2011)]{navarro2011} Navarro, J.~F., Abadi, 
M.~G., Venn, K.~A., Freeman, K.~C., 
\& Anguiano, B.\ 2011, \mnras, 412, 1203 

\bibitem[Nicolet(1981)]{nicolet1981} Nicolet, B.\ 1981, \aap, 104, 185 

\bibitem[Nissen(1988)]{nissen1988} Nissen, P.~E.\ 1988, \aap, 199, 146 
\bibitem[Nordstr{\"o}m et al.(2004)]{nordstrom2004} Nordstr{\"o}m, B., et al.\ 2004, \aap, 418, 989 

\bibitem[Origlia et al.(2006)]{origlia2006} Origlia, L., Valenti,  E., Rich, R.~M., \&
Ferraro, F.~R.\ 2006, \apj, 646, 499 

\bibitem[Pace et  al.(2008)]{pace2008} Pace, G., Pasquini, L., \& Fran{\c c}ois, P.\ 2008, \aap, 489, 403 

\bibitem[Pace et al.(2010)]{pace2010} Pace, G., Danziger, J., Carraro, G., Melendez, J., Fran{\c c}ois, P., Matteucci, F., \& Santos, N.~C.\ 2010, \aap, 515, A28 

\bibitem[Pancino et al.(2010a)]{pancino2010a} Pancino, E., Carrera, R., Rossetti, E., \& Gallart, C.\ 2010, \aap, 511, 56. Paper I

\bibitem[Pancino et  al.(2010b)]{pancino2010b} Pancino, E., Rejkuba, M., Zoccali,
M., \& Carrera, R.\ 2010, \aap, 524, A44 

\bibitem[Pasquini et  al.(2004)]{pasquini2004} Pasquini, L., Randich, S., Zoccali, M.,
Hill, V., Charbonnel, C., \& Nordstr{\"o}m, B.\ 2004, \aap, 424, 951 

\bibitem[Patenaude(1978)]{patenaude1978} Patenaude, M.\ 1978, \aap, 66, 225 

\bibitem[Paulson et al.(2003)]{paulson2003} Paulson, D.~B., Sneden, 
C., \& Cochran, W.~D.\ 2003, \aj, 125, 3185 

\bibitem[Pereira \& Quireza(2010)]{pereira2010} Pereira, C.~B., \& Quireza, C.\ 2010, IAU Symposium, 266, 495 

\bibitem[Percival et al.(2003)]{percival2003} Percival, S.~M., Salaris, M., \& Kilkenny, D.\ 2003, \aap, 400, 541 

\bibitem[Perryman et al.(1997)]{perryman1997} Perryman, M.~A.~C., et al.\ 1997, \aap, 323, L49 

\bibitem[Perryman et al.(1998)]{perryman1998} Perryman, M.~A.~C., et al.\ 1998, \aap, 331, 81

\bibitem[Peterson \& Green(1998)]{peterson1998} Peterson, R.~C., \& Green, E.~M.\ 1998,
\apjl, 502, L39 

\bibitem[Pfeiffer et al.(1998)]{foces} Pfeiffer, M.~J., Frank, C., Baumueller, D., Fuhrmann, K., \& Gehren, T.\ 1998, \aaps, 130, 381 

\bibitem[Piatti et al.(1995)]{piatti1995} Piatti, A.~E., Claria, J.~J., \& Abadi, M.~G.\ 1995, \aj, 110, 2813 


\bibitem[Pinsonneault et al.(1998)]{pinsonneault1998} Pinsonneault, M.~H., Stauffer, J., Soderblom, D.~R., King, J.~R., 
\& Hanson, R.~B.\ 1998, \apj, 504, 170 

\bibitem[Platais et al.(2008)]{platais2008} Platais, I., Melo, C.,  Fulbright, J.~P., Kozhurina-Platais, V., Figueira, P., Barnes, S.~A., \& M{\'e}ndez, R.~A.\ 2008, \mnras, 391, 1482 


\bibitem[Pourbaix et al.(2004)]{pourbaix2004} Pourbaix, D., et al.\ 2004, \aap, 424, 727 
\bibitem[Ram{\'{\i}}rez \& Cohen(2003)]{ramirez2003}Ram{\'{\i}}rez, S.~V.,\& Cohen, J.~G.\ 2003, \aj, 125, 224 

\bibitem[Randich et al.(2001)]{randich2001} Randich, S., Pallavicini, R., Meola, G.,
Stauffer, J.~R., \& Balachandran, S.~C.\ 2001, \aap, 372, 862 

\bibitem[Randich et al.(2003)]{randich2003} Randich, S., Sestito, P., \& Pallavicini,
R.\ 2003, \aap, 399, 133 

\bibitem[Randich et al.(2006)]{randich2006} Randich, S., Sestito, P., Primas, F.,
Pallavicini, R., \& Pasquini, L.\ 2006, \aap, 450, 557 

\bibitem[Reddy et al.(2003)]{reddy2003} Reddy, B.~E., Tomkin, J.,  Lambert, D.~L.,
\& Allende Prieto, C.\ 2003, \mnras, 340, 304 

\bibitem[Reddy et al.(2006)]{reddy2006} Reddy, B.~E., Lambert,  D.~L., \& Allende
Prieto, C.\ 2006, \mnras, 367, 1329 

\bibitem[Reid et al.(2007)]{reid2007} Reid, I.~N., Turner, E.~L., Turnbull, M.~C., Mountain, M., \& Valenti, J.~A.\ 2007, \apj, 665, 767 

\bibitem[Richtler \& Sagar(2001)]{Richtler2001} Richtler, T., \& Sagar, R.\ 2001, Bulletin of the Astronomical Society of India, 29, 53 

\bibitem[Roman(1955)]{roman1955} Roman, N.~G.\ 1955, \apj, 121, 
454 
\bibitem[Rohlfs \& Vanysek(1962)]{rohlfs1962} Rohlfs, K., \& Vanysek, V.\ 1962, Astronomische Abhandlungen der Hamburger Sternwarte, 5, 341 

\bibitem[Ro{\v s}kar et al.(2008)]{roskar2008} Ro{\v s}kar, R., Debattista, V.~P., Quinn, T.~R., Stinson, G.~S., \& Wadsley, J.\ 2008, \apjl, 684, L79 


\bibitem[Salaris et al.(2004)]{salaris2004} Salaris, M., Weiss, A., \& Percival, S.~M.\ 2004, \aap, 414, 163

\bibitem[Santos et al.(2009)]{santos2009} Santos, N.~C., Lovis, C., Pace, G., Melendez, J., \& Naef, D.\ 2009, \aap, 493, 309 

\bibitem[Schuler et al.(2003)]{schuler2003} Schuler, S.~C., King,  J.~R., Fischer,
D.~A., Soderblom, D.~R.,  \& Jones, B.~F.\ 2003, \aj, 125, 2085 

\bibitem[Schuler et al.(2006)]{schuler2006} Schuler, S.~C., Hatzes, A.~P., King, J.~R., K{\"u}rster, M., \& The, L.-S.\ 2006, \aj, 131, 1057 

\bibitem[Schuler et al.(2009)]{schuler2009} Schuler, S.~C., King, J.~R., \& The, L.-S.\ 2009, \apj, 701, 837 

\bibitem[Sestito et  al.(2003)]{sestito2003} Sestito, P., Randich, S., Mermilliod,
J.-C., \& Pallavicini, R.\ 2003, \aap, 407, 289 

\bibitem[Sestito et al.(2004)]{sestito2004} Sestito, P., Randich, S., \& Pallavicini, R.\ 2004, \aap, 426, 809 

\bibitem[Sestito et al.(2006)]{sestito2006} Sestito, P., Bragaglia, A., Randich, S., Carretta, E., Prisinzano, L., \& Tosi, M.\ 2006, \aap, 458, 121

\bibitem[Sestito et al.(2007)]{sestito2007} Sestito, P., Randich, S., \& Bragaglia, A.\
2007, \aap, 465, 185 

\bibitem[Sestito et al.(2008)]{sestito2008} Sestito, P., Bragaglia, A., Randich, S.,
Pallavicini, R., Andrievsky, S.~M., \& Korotin, S.~A.\ 2008, \aap, 488, 943 

\bibitem[Schaye(2004)]{schaye2004} Schaye, J.\ 2004, \apj, 609, 
667 

\bibitem[Shen et al.(2005)]{shen2005} Shen, Z.-X., Jones, B., 
Lin, D.~N.~C., Liu, X.-W., \& Li, S.-L.\ 2005, \apj, 635, 608 

\bibitem[Skrutskie et al.(2006)]{2mass2} Skrutskie, M.~F., et  al.\ 2006, \aj, 131, 1163

\bibitem[Smiljanic et al.(2009)]{smiljanic2009} Smiljanic, R., Gauderon, R., North, P., Barbuy, B., Charbonnel, C., \& Mowlavi, N.\ 2009, \aap, 502, 267 

\bibitem[Smith \& Suntzeff(1987)]{smith1987} Smith, V.~V., \& Suntzeff, N.~B.\ 1987,
\aj, 93, 359

\bibitem[Soderblom et al.(2009)]{soderblom2009} Soderblom, D.~R., 
Laskar, T., Valenti, J.~A., Stauffer, J.~R., 
\& Rebull, L.~M.\ 2009, \aj, 138, 1292 

\bibitem[Soubiran \& Girard(2005)]{soubiran2005} Soubiran, C., \& Girard, P.\ 2005, \aap, 438, 139 

\bibitem[Soubiran et al.(2008)]{soubiran2008} Soubiran, C., Bienaym{\'e}, O., Mishenina, T.~V., \& Kovtyukh, V.~V.\ 2008, \aap, 480, 91 

\bibitem[Spite(1967)]{spite1967} Spite, M.\ 1967, Annales 
d'Astrophysique, 30, 211 

\bibitem[Stanghellini \& Haywood(2010)]{stanghellini2010} Stanghellini, L., \& Haywood, M.\ 2010, \apj, 714, 1096 

\bibitem[Stetson \& Pancino(2008)]{daospec} Stetson, P.~B., \& Pancino, E.\ 2008, \pasp, 120, 1332

\bibitem[Tadross(2001)]{tadross2001} Tadross, A.~L.\ 2001, New Astronomy, 6, 293 

\bibitem[Tautvai\^siene et al.(2000)]{tautvaisiene2000} Tautvai\^siene, G., Edvardsson, B.,
Tuominen, I., Ilyin, I.\ 2000, A\&A, 360, 499

\bibitem[Tautvai{\v s}ien{\.e} et  al.(2005)]{tautvaisiene2005} Tautvai{\v s}ien{\.e}, G.,
Edvardsson, B., Puzeras, E., \& Ilyin, I.\ 2005, \aap, 431, 933 

\bibitem[Taylor(2006)]{taylor2006} Taylor, B.~J.\ 2006, \aj, 132, 2453 

\bibitem[Taylor(2007)]{taylor2007} Taylor, B.~J.\ 2007, \aj, 134, 934 

\bibitem[Terndrup et al.(2002)]{terndrup2002} Terndrup, D.~M., 
Pinsonneault, M., Jeffries, R.~D., Ford, A., Stauffer, J.~R., 
\& Sills, A.\ 2002, \apj, 576, 950 

\bibitem[Tosi et al.(2007)]{tosi2007} Tosi, M., Bragaglia, A., \& Cignoni, M.\ 2007, \mnras, 378, 730 

\bibitem[Tsvetkov(1993)]{tsvetkov1993} Tsvetkov, T.~G.\ 1993, \apss, 203, 247 


\bibitem[Twarog(1983)]{twarog1983} Twarog, B.~A.\ 1983, \apj, 267, 207 

\bibitem[Twarog et al.(1997)]{twarog1997} Twarog, B.~A., Ashman, K.~M., \& Anthony-Twarog, B.~J.\ 1997, AJ, 114, 2556 

\bibitem[Twarog et al.(2003)]{twarog2003} Twarog, B.~A., Anthony-Twarog, B.~J., \& De Lee, N.\ 2003, \aj, 125, 1383 

\bibitem[van Bueren(1952)]{vanBueren1952} van Bueren, H.~G.\ 1952, \bain, 11, 385

\bibitem[van den Heuvel(1969)]{vandenheuvel1969} van den Heuvel, 
E.~P.~J.\ 1969, \pasp, 81, 815 

\bibitem[van Leeuwen(1999)]{vanleeuwen1999} van Leeuwen, F.\ 1999, \aap, 341, L71 

\bibitem[Varenne \& Monier(1999)]{varenne1999} Varenne, O., \& Monier, R.\ 1999, \aap, 351, 247 

\bibitem[Villanova et al.(2005)]{villanova2005} Villanova, S.,  Carraro, G., Bresolin,
F., \& Patat, F.\ 2005, \aj, 130, 652 

\bibitem[Villanova et al.(2007)]{villanova2007} Villanova, S., Baume,  G., \& Carraro,
G.\ 2007, \mnras, 379, 1089 

\bibitem[Villanova et al.(2009)]{villanova2009} Villanova, S., Carraro, G., \& Saviane, I.\ 2009, \aap, 504, 845 

\bibitem[Villanova et al.(2010)]{villanova2010} Villanova, S., Geisler, D., \& Piotto, G.\ 2010, \apjl, 722, L18 


\bibitem[Yong et al.(2005)]{yong2005}Yong D., Carney B.W., \& Texeira de
Almeida M.L.\ 2005, AJ, 130, 597

\end{thebibliography}
\end{document}